\documentclass[%
reprint,
amsmath,amssymb,
aps,
]{revtex4-1}

\usepackage{graphicx}
\usepackage{dcolumn}
\usepackage{bm}
\usepackage{bbm}
\usepackage{mathrsfs}

\usepackage[table]{xcolor}
\usepackage{ulem} 
\DeclareMathOperator{\sign}{sign}
\begin{document}
	
	\preprint{APS/123-QED}
	
	\title{\large\bf Effect of local Coulomb interaction on Majorana corner modes: \\ weak and strong correlation limits}
	
	
\author{S.\,V.\, Aksenov}%
\email{asv86@iph.krasn.ru}
\author{A.\,D.\, Fedoseev}%
\email{fad@iph.krasn.ru}
\author{M.\,S.\, Shustin}%
\email{mshustin@yandex.ru}
\author{A.\,O.\, Zlotnikov}
\email{zlotn@iph.krasn.ru}

	\affiliation{%
		Kirensky Institute of Physics, Federal Research Center KSC SB RAS, 660036 Krasnoyarsk, Russia}
	
	\date{\today}
	
	\begin{abstract}

	Here we present an analysis of the evolution of Majorana corner modes realizing in a higher-order topological superconductor (HOTSC) on a square lattice under the influence of local Coulomb repulsion. The HOTSC spectral properties were considered in two regimes: when the intensities of many-body interactions are either weak or strong. The weak regime was studied using the mean-field approximation with self-consistent solutions carried out both in the uniform case and taking into account of the boundary of the finite square-shaped system. It is shown that in the uniform case the topologically nontrivial phase on the phase diagram is widened by the Coulomb repulsion. The boundary effect, resulting in an inhomogeneous spatial distribution of the correlators, leads to the appearance of the crossover from the symmetric spin-independent solution to the spin-dependent one characterized by a spontaneously broken symmetry. In the former the corner states have energies that are determined by the overlap of the excitation wave functions localized at the different corners. In the latter the corner excitation energy is defined by the Coulomb repulsion intensity with a quadratic law. The crossover is a finite size effect, i.e. the larger the system the lesser the critical value of the Coulomb repulsion. In the strong repulsion regime we derive the effective HOTSC Hamiltonian in the atomic representation and found a rich variety of interactions induced by virtual processes between the lower and upper Hubbard subbands. It is shown that Majorana corner modes still can be realized in the limit of the infinite repulsion. Although the boundaries of the topologically nontrivial phase are strongly renormalized by Hubbard corrections.

\begin{description}
	\item[PACS number(s)]
	71.10.Pm, 
	74.78.Na, 
	74.45.+c, 
\end{description}

\end{abstract}	
		
\maketitle


\section{\label{sec1}Introduction}

The development of the concept of topologically nontrivial systems has led in recent years to an active study of high-order topological insulators and superconductors (HOTSCs) \cite{benalcazar-17,Langbehn-17,zlotnikov-21}. The spectrum of their both bulk and edge states has a gap. In turn, topologically protected gapless excitations arise, being localized at the boundaries of higher orders, i.e. at corners (corners and hinges) in 2D (3D) systems \cite{volovik-10}. It is important to note that in case of 2D HOTSCs such states are Majorana corner modes (MCMs) which possess zero energy and obey non-Abelian exchange statistics \cite{ivanov-01,alicea-11}.

Taking into account ongoing attempts to utilize Majorana modes for the realization of quantum computations, their "corner species" have a natural advantage over the Majoranas emerging in 1D systems \cite{kitaev-01,lutchyn-10,oreg-10}. The latter require a purely 1D system, while the finite width of the wire leads to the appearance of a gapless band of edge excitations. In this case, the zero-energy Majoranas, still detached from bulk states by a gap, are no longer separated from other edge excitations. In addition, as the 1D system is widened, the character of the excitations changes from purely Majorana to chiral \cite{Potter-10,Sedlmayr-16} with a change in the ratio between the length and width of the system. Moreover, the braiding procedure (the spatial exchange of the Majorana modes resulting in the phase shift of the ground state wave function) can only be carried out in 2D system \cite{Nayak-08}, so one need to construct 2D devices from 1D topological superconductors \cite{Cheng-16,Harper-19,Zhou-20} to achieve this goal.

The predicted MCMs solve these problems. First, their energy lies in the gap of the spectrum of both bulk and edge excitations. Secondly, their localization strictly in the corners of the system prevents their Majorana character from changing regardless of the size ratio of the system. Additional interest in HOTSCs is caused by the possibility to move the corner excitations by varying the parameters of the system. In particular, in a number of works a magnetic field is used to create HOTSC \cite{Zhu-18,Franca-19,Wu-20,Plekhanov-21}. It plays the role of a perturbation destroying the symmetry that underlies the first-order topological system. In some cases, the MCM position can be controlled using the direction of this magnetic field \cite{pahomi-20,zhang-20PRR}. A model including triangular HOTSC segments has also recently been proposed demonstrating the possibility of braiding using only electric fields \cite{zhang-20PRB}. Thus, the MCMs in 2D systems seem to be good candidates for braiding, which is one of the key requirements for creating a topological qubit. Another possible practical application of such systems that deserves attention is conventional nanoscale devices with controlled transport characteristics.

Despite the active study of HOTSCs, there are still many unresolved issues. First, the influence of Coulomb correlations on the conditions of the topological phase transition and MCMs properties remains poorly understood. There are works in which superconducting pairing, which generates the corner states, is calculated self-consistently, taking into account the Coulomb interaction in the system \cite{hsu-20,kheirkhah-20,li-22}. However, many of the previously proposed models imply the introduction of superconducting pairing due to the proximity effect. The question of how the obtained results would change if there are Coulomb correlations in the system itself is not fully resolved yet. At the same time, it is known that taking the local repulsion into account can significantly affect the properties of conventional topological superconductors \cite{stoudenmire-11,thomale-13,katsura-15,aksenov-20}. In the case of higher-order topological insulators, the many-body interactions can lead both to the appearance of new topological classes \cite{kudo-19,otsuka-21} and, conversely, to the destruction of topological states in 3D systems \cite{zhao-21}.

Secondly, while higher-order 2D topological phases have already been experimentally demonstrated in photonic, acoustic and topoelectric systems \cite{el_hassan-19,ni-19,imhof-18,serra-garcia-19}, their solid-state counterparts have not been realized yet. Moreover, bismuth is the only material confirmed to provide the higher-order topology \cite{Schindler-18NP,Aggarwal-21}, although some uncertainty still remains \cite{Drozdov-14}. Other HOTI and HOTSC candidates are transition-metal dichalcogenides \cite{Wang-19TMD,Ezawa-19TMD,Qian-22} and rocksalt IV–VI semiconductors XY (X = Ge, Sn, Pb and Y = S, Se, Te) \cite{Wrasse-14,Liu-15}, but their higher-order topology has not been confirmed experimentally yet. Remarkably, it has been already found out that spectral and transport properties of some of these 2D topological insulators can significantly depend on electron-electron interactions \cite{Sante-17,Sihi-21,Sihi-22}. Thus, the problem of the local Coulomb (Hubbard-type) repulsion in 2D solid-state HOTSC is of fundamental nature and its solution will make it possible to better estimate the prospects for the experimental detection of the MCMs.

The present article is devoted to the study of the Hubbard interaction problem in a typical HOTSC model. We analyze both limits of weak and strong repulsion. Based on this, the rest of article is organized as follows. In Sec. \ref{sec2} we describe a HOTSC Hamiltonian. The effect of weak intraorbital Coulomb repulsion on the MCMs is discussed in Sec. \ref{sec3}. In Sec. \ref{sec4} we analyze an effective Hamiltonian of strongly-correlated HOTSC and its topological features. We conclude in Sec. \ref{sec5} with a summary. In Appendix \ref{apxA} the conditions of the HOTSC phase realization are obtained employing an effective mass criterion. We discuss the derivation of an effective Hamiltonian in the regime of the strong finite Hubbard interaction in Appendix \ref{apxB}. Appendix \ref{apxC} deals with a Green functions approach in the $U\to\infty$ limit.

\section{\label{sec2} Model Hamiltonian}

One of the criteria used to describe the higher-order topological phase transition is a so-called change of effective mass sign. It's known that the MCMs arise if two initially gapless topological
states propagating along the adjacent edges acquire an effective mass of the opposite sign due to an interaction that breaks one of the symmetries responsible for the first-order nontrivial topology. In this situation the corner can be treated as a domain wall or, in other words, as a topological defect. Below we describe one of the popular 2D models possessing this feature and used to study physics of the MCMs on a square lattice. In order to obtain the gap in the edge spectrum induced by some interaction it is necessary to initially prepare two subsystems with inverted bands. One of the proper candidates is a bipartite square lattice with an interorbital Rashba spin-orbit coupling where an extended $s$($d$)-wave intraorbital pairing plays a role of the interaction \cite{Wang-18}. The corresponding tight-binding Hamiltonian is
\begin{eqnarray} \label{H01}
    &&H_{0} =\sum_{f\eta\sigma}\left(\eta\Delta\varepsilon-\mu\right)c^{+}_{f\eta\sigma}c_{f\eta\sigma}\\
    &&+\sum_{\eta}\eta\left( \sum_{\langle fm\rangle_{x},\sigma}t_{x}+\sum_{\langle fm\rangle_{y},\sigma}t_{y}+\sum_{\langle\langle fm\rangle\rangle,\sigma}t_{1}\right)c^{+}_{f\eta\sigma}c_{m\eta\sigma}\nonumber\\
    &&+i\alpha\sum_{\langle fm\rangle}\left[\hat{\tau}^{\alpha\beta},e_{fm}\right]_{z}\hat{\sigma}^{\nu\eta}_{x}c^{+}_{f\nu\alpha}c_{m\eta\beta}\nonumber\\
    &&+\left(\Delta_{x}\sum_{\langle fm\rangle_x,\eta} + \Delta_{y}\sum_{\langle fm\rangle_y,\eta}\right)c^{+}_{f\eta\uparrow}c^{+}_{m\eta\downarrow}\nonumber\\
    &&+\Delta_{0}\sum_{f\eta}c^{+}_{f\eta\uparrow}c^{+}_{f\eta\downarrow}+h.c.,\nonumber
\end{eqnarray}
where $c_{f\eta\sigma}$ annihilates an electron with a spin $\sigma$ on an $\eta$th orbital ($\eta=A,B$) at a square lattice site $f=\left(i,j\right)$; $i,j=1,...,N$; $\Delta\varepsilon$ is an on-site energy shift opposite for different orbitals; $\mu$ is a chemical potential. The intraorbital nearest-neighbor $t_{x,y}$ as well as next-nearest-neighbor $t_{1}$ hopping parameters are of opposite signs for different orbitals leading to the inverted bands. The parameter $\alpha$ defines an intensity of the interorbital Rashba spin-orbit coupling; $e_{fm}$ is a unit vector pointing along the direction of electron motion from the $m$th to $f$th site. The parameters $\Delta_{0,x,y}$ are intensities of the intraorbital on-site and intersite singlet pairing that results in overall $s_{\pm}$-wave superconductivity in the case $\Delta_x=\Delta_y$ or $s+d_{x^2-y^2}$-wave superconductivity in the case of $\Delta_x=-\Delta_y$. Unless otherwise specified, it will be assumed that $\Delta_x=\Delta_y=\Delta_{1}$. The Pauli matrices $\hat{\sigma}_{n}$ and $\hat{\tau}_{n}$ ($n=x,y,z$) act in orbital and spin subspaces, respectively.

The goal of this study is to analyze the effects of local intraorbital Coulomb repulsion on the topological properties and corner excitations of the model \eqref{H01}. Then, the total Hamiltonian is
\begin{equation} \label{H}
    H=H_{0}+H_{U}.
\end{equation}
The last term in \eqref{H} is responsible for the many-body interactions read
\begin{equation} \label{HU}
H_{U} =\sum\limits_{f\eta}U_{\eta}n_{f\eta\uparrow}n_{f\eta\downarrow},
\end{equation}
where
$U_{\eta=A,B}$ - a strength of the intraorbital Coulomb interaction; $n_{f\eta\sigma}$ is an orbital-dependent electron number operator at the site $f$. In the subsequent Sections our attention will be drawn to the two limits of weak and strong charge correlations. For the sake of simplicity, it will be assumed there that $U_{A}=U_{B}=U$.

\section{\label{sec3} Weak Coulomb interaction}

\subsection{\label{sec3.1} Mean-field approximation for the two-orbital HOTSC Hamiltonian}

We start the analysis of the problem with the regime of the weak Coulomb repulsion. Here one can employ the usual mean-field approximation to reduce the Hamiltonian \eqref{HU} to a quadratic form, the spectral properties of which, in turn, can be found using the Bogolyubov transformation. Technically, in this case the summand \eqref{HU} is reduced to

\begin{eqnarray} \label{HU_weak}
&&H^{w}_{U} \approx U\sum\limits_{f\eta\sigma}\left[\langle n_{f\eta\sigma}\rangle n_{f\eta\bar{\sigma}}-\langle c_{f\eta\sigma}^{+}c_{f\eta\bar{\sigma}}\rangle c_{f\eta\bar{\sigma}}^{+}c_{f\eta\sigma}\right]-\\
&&~~~~-U\sum\limits_{f\eta}\left[\langle c_{f\eta\uparrow}^{+}c_{f\eta\downarrow}^{+}\rangle c_{f\eta\uparrow}c_{f\eta\downarrow}-\langle c_{f\eta\uparrow}c_{f\eta\downarrow}\rangle c_{f\eta\uparrow}^{+}c_{f\eta\downarrow}^{+}\right].\nonumber
\end{eqnarray}
Thus, intraorbital Hubbard interaction results in corrections of the on-site particle energies which are proportional to the average occupations. Next, the on-site spin-flip terms arise that, in general, can be interpreted as an influence of longitudinal magnetic field. The last two terms in \eqref{HU_weak} give the corrections to the on-site singlet pairing amplitude $\Delta_{0}$.

The averages in \eqref{HU_weak} can be found in a standard manner using the Bogolyubov $u,v$-coefficients,
\begin{eqnarray}
\label{corrs_weak}
&&\langle c^{+}_{f\eta\sigma}c_{f\eta'\sigma'} \rangle=\sum_{n=1}^{4N^2}\Big[ u_{f\eta n\sigma}u^{*}_{f\eta'n\sigma'}f\left(\frac{\varepsilon_n}{T}\right) + \Big.\\
&&\Big.~~~~~~~~~~~~~~~~~~~~~~~~~~~~~~~~~~ + v_{f\eta n\sigma}v^{*}_{f\eta'n\sigma'}\left( 1 - f\left(\frac{\varepsilon_n}{T}\right) \right)\Big],\nonumber\\
&&\langle c^{+}_{f\eta\sigma}c^{+}_{f\eta'\sigma'} \rangle=\sum_{n=1}^{4N^2}\Big[ u_{f\eta n\sigma}v^{*}_{f\eta'n\sigma'}f\left(\frac{\varepsilon_n}{T}\right) + \Big. \nonumber\\
&& \Big.~~~~~~~~~~~~~~~~~~~~~~~~~~~~~~~~~~ + v_{f\eta n\sigma}u^{*}_{f\eta'n\sigma'}\left( 1 - f\left(\frac{\varepsilon_n}{T}\right) \right)\Big],\nonumber
\end{eqnarray}
where $f(\varepsilon_n/T)$ - the Fermi-Dirac distribution function of the $n$th Bogolyubov excitation with an energy $\varepsilon_n$ and $(u,v)_{f\eta n\sigma}$ are corresponding coefficients. Then, the self-consistent calculation of the spectrum of $H^{w}=H_{0}+H^{w}_{U}$ and correlators \eqref{corrs_weak} allows to analyze the influence of the weak local Coulomb repulsion on the MCMs.

\subsection{\label{sec3.2} Coulomb interaction effect on the HOTSC in the uniform case}

\begin{figure}[h!]
\begin{center}
\includegraphics[width=0.5\textwidth]{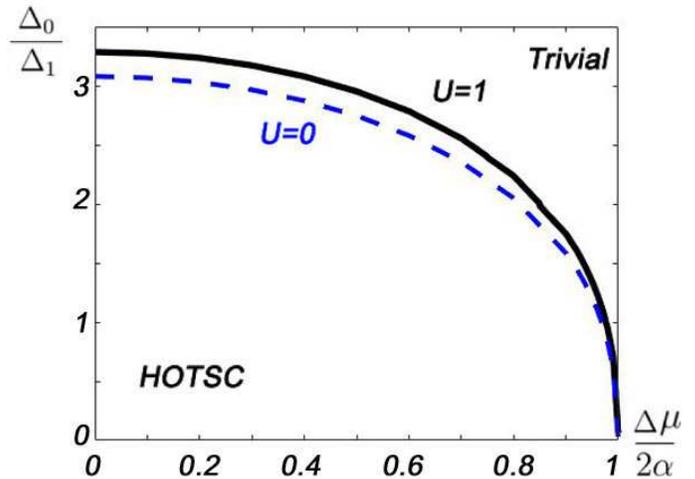}
\caption{Topological phase diagram of the 2D square-shaped topological insulator with extended $s$-wave superconducting coupling without Coulomb interaction $U=0$ (blue dashed line, according to (\ref{HOTSC_cond1})) and with on-site Coulomb interaction $U=1$ (black solid line). $\Delta\mu$ is chemical potential measured from the half-filling level. The other parameters are $\Delta\varepsilon=0$, $t_x=-t_y=2$, $t_1=t_x/2$, $\Delta_{1}=0.5$, $\alpha=1.5$.}\label{proc_fig}
\end{center}
\end{figure}

We start our analysis of Coulomb interaction effect on the topological properties of HOTSC with uniform case in the $T=0$ limit. In this situation the correlators included in (\ref{HU_weak}) supposed to be independent of the site number and the impact of the boundary on them is neglected. The correlators are calculated self-consistently under the periodic boundary conditions.

The numerical investigation shows that the influence of the intraorbital Coulomb interaction in such a case reduces to corrections of the on-site energies and corresponding singlet superconducting coupling. The former implies the modification of $\Delta\varepsilon\rightarrow\widetilde{\Delta\varepsilon}$ parameter and the shift of both bands, which do not affect the topological properties of the system. The second correction is a well known suppression of the on-site superconducting coupling $\Delta_0\rightarrow\widetilde{\Delta}_0$. Thus the topological properties of the system remain qualitatively the same up to the modification of $\Delta\varepsilon$ and $\Delta_0$ parameters. Quantitatively, the change of $\Delta\varepsilon$ is small and its effect on the topological phase diagram is insignificant compared with the $\Delta_0$ correction. As the on-site singlet coupling suppresses the higher-order topological phase (see \cite{Wang-18} for qualitative explanation and Appendix \ref{apxA} for mathematical details), its reduction with the $U$ increase stabilizes the nontrivial phase and widens the corresponding region on the topological diagram (Fig. \ref{proc_fig}).

\subsection{\label{sec3.3} Self-consistent solution in the open boundary conditions case}

Now we proceed with the case of square-shaped HOTSC with open boundary conditions. In such situation the correlators (\ref{corrs_weak}) become dependent on the site index leading to the inability of topological phase analysis. Meanwhile, the properties of the corner excitations still can be investigated.

\begin{figure}[!htb]
    \includegraphics[width=0.5\textwidth]{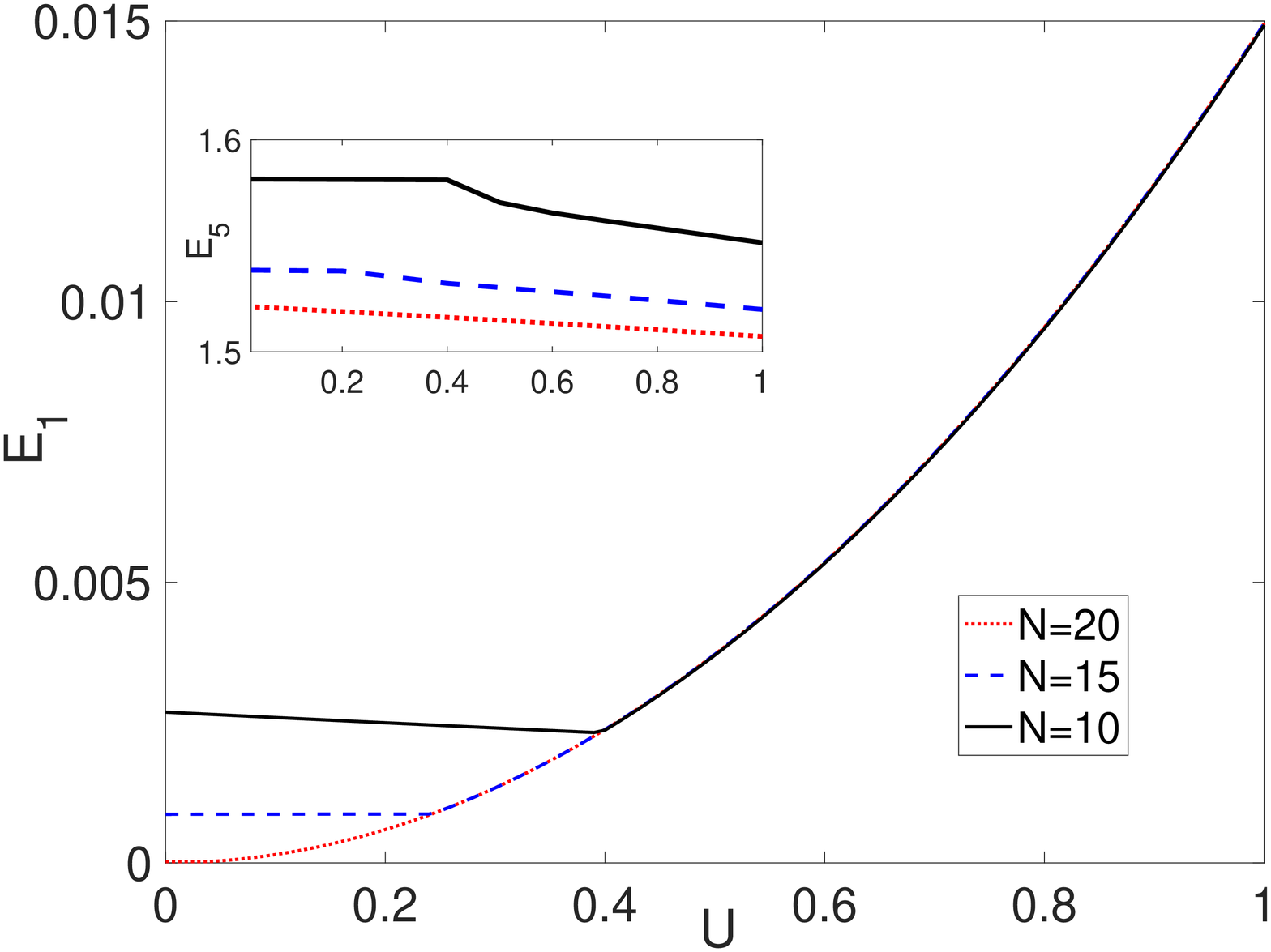}
    \caption{\label{1} Dependence of the first excitation energy on the intensity of the intraorbital Hubbard repulsion, $E_{1}\left(U\right)$, for different sizes of the system. Inset: the energy of the first out-of-gap state as a function of $U$. The system is taken at half filling ($\mu=U/2$) with $\Delta_{0}=0$. The other parameters are the same as in Fig. \ref{proc_fig}.}
\end{figure}

We carried out series of self-consistent calculations for different parameters of the model. The typical dependence of the first excitation energy on the intensity of the intraorbital Hubbard repulsion, $\varepsilon_{1}\left(U\right)$, for different sizes of the system is plotted in Figure \ref{1}. The numerical calculations revealed the presence of crossover between two qualitatively different cases. For $U<U_c$ the corner excitations remain almost unperturbed by the Coulomb repulsion with their energies being determined by the overlapping of excitations in different corners of the finite-size system. For $U>U_c$ the energies depend quadratically on $U$. Note that there is still a considerable gap in the spectrum of the open system between the corner states ($E_{n=1-4}$) and the rest of the excitations even at $U=t_{x}/2$ (see the inset of Fig. \ref{1}a).

To understand the qualitative difference between the solutions before and after the crossover, it is necessary to analyze the correlators \eqref{corrs_weak}. Since time-reversal symmetry is preserved in the bare Hamiltonian $H_{0}$ the self-consisted calculation at $U\neq0$ does not generate the nonzero normal spin-flip averages, $\langle c^{+}_{f\eta\sigma}c_{f\eta\bar{\sigma}} \rangle=0$. Then, the block-diagonal structure of the system Bogolyubov-de-Gennes Hamiltonian in the basis $\left[c_{fA\sigma},~c_{fB\bar{\sigma}},~c^{+}_{fA\bar{\sigma}},~c^{+}_{fB\sigma} \right]$ remains. Taking it into account, it is convenient to consider the corresponding sums of the on-site concentration averages, $\langle n_{fA\sigma} \rangle + \langle n_{fB\bar{\sigma}} \rangle$, as they describe possible spatial fluctuations relative to the quarter filling, which are induced by the Hubbard repulsion.

\begin{figure*}[!htb]
    \includegraphics[width=0.45\textwidth]{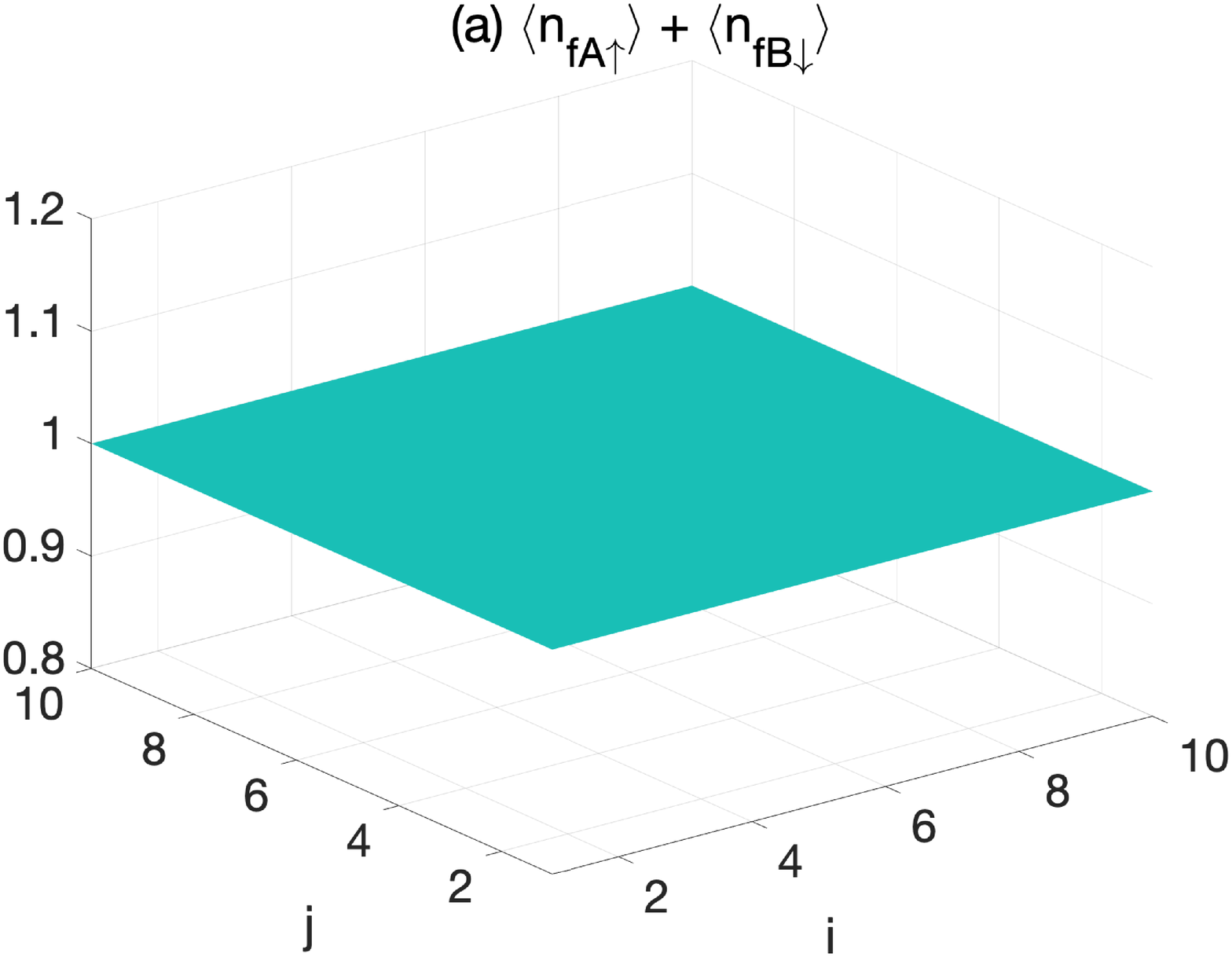}
    \includegraphics[width=0.45\textwidth]{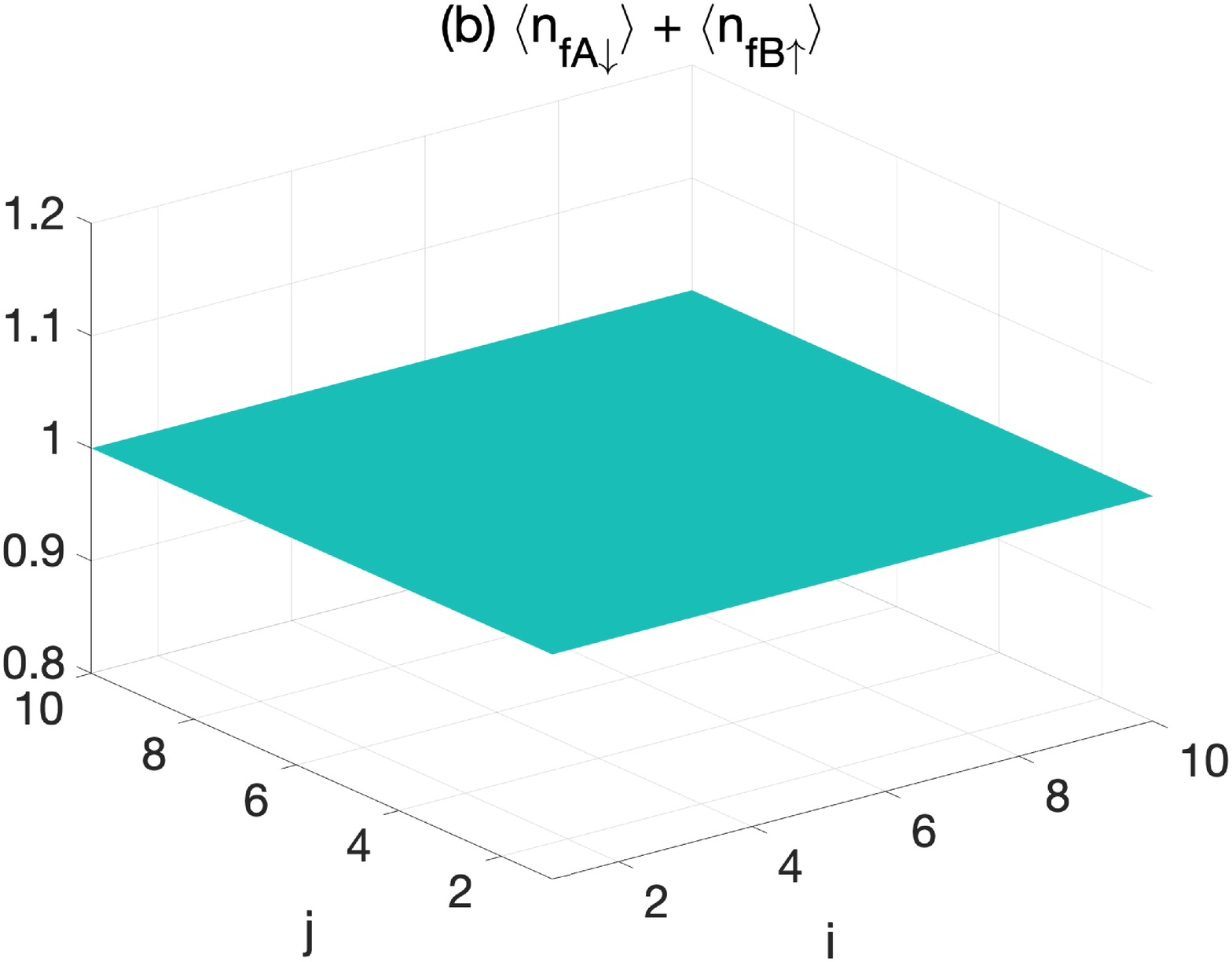}
    \includegraphics[width=0.45\textwidth]{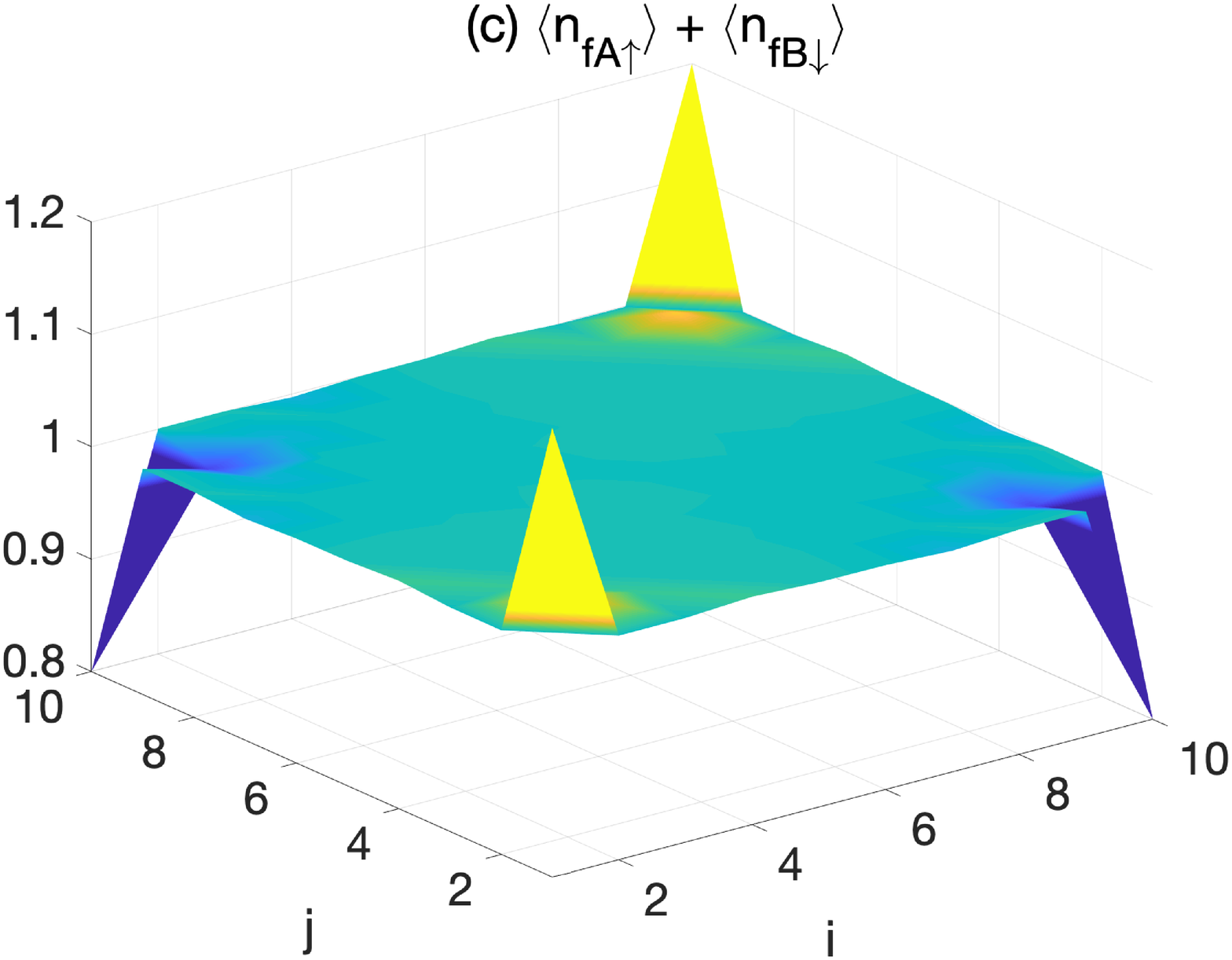}
    \includegraphics[width=0.45\textwidth]{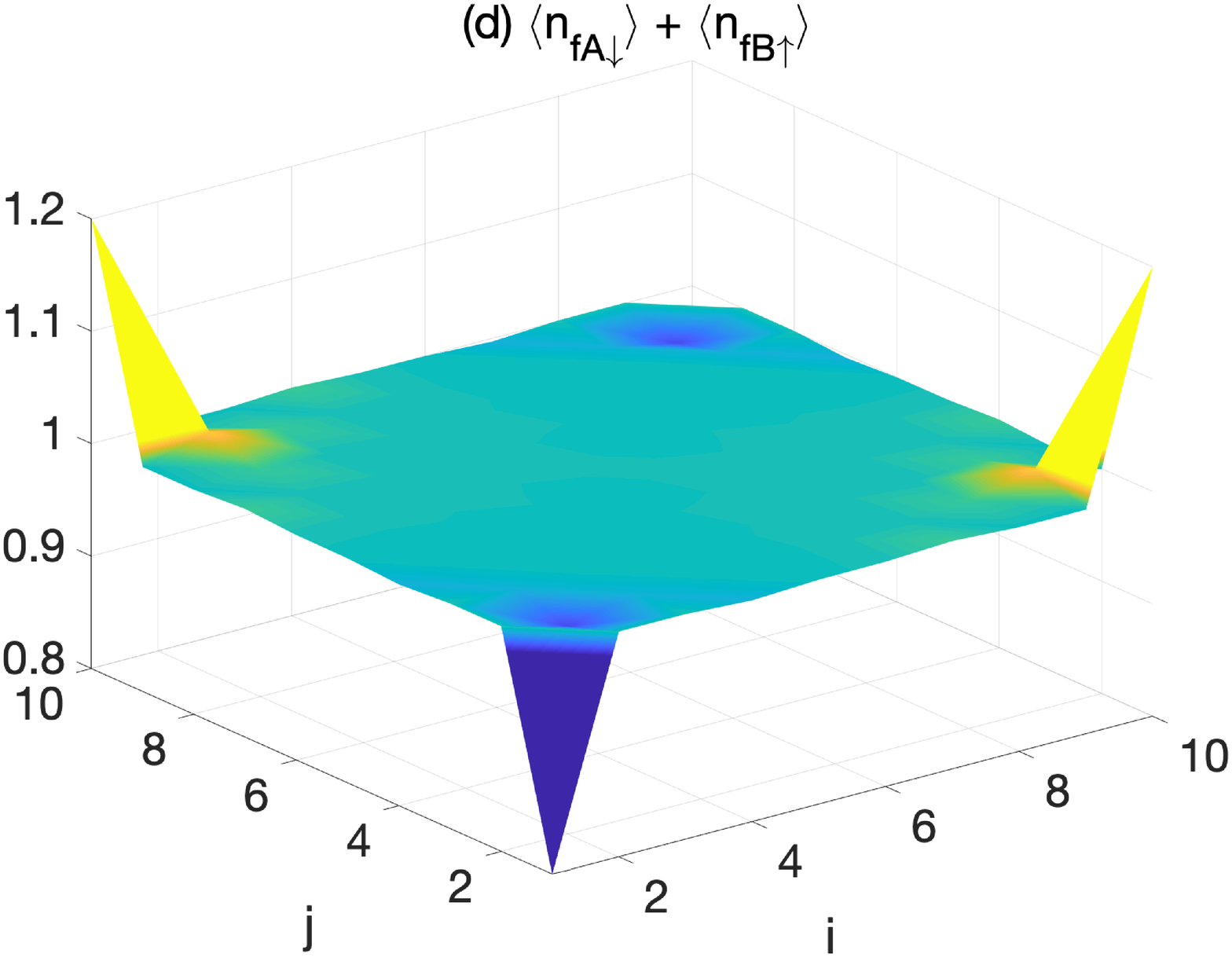}
    \caption{\label{2} Spatial distribution of the correlators $\langle n_{fA\uparrow} \rangle + \langle n_{fB\downarrow} \rangle$ and $\langle n_{fA\downarrow} \rangle + \langle n_{fB\uparrow} \rangle$ in the $C_{4}$-symmetric phase (a,b) and in the phase with the spontaneously broken $C_{4}$ symmetry (c,d). Parameters: $N=10$. }
\end{figure*}

The dependencies $\langle n_{fA\sigma} \rangle + \langle n_{fB\bar{\sigma}} \rangle$ at $U<U_{c}$ are displayed in Figs. \ref{2}a and \ref{2}b. One can note that in both half-spaces $C_{4}$ symmetry persists. Additionally, the separate distributions $\langle n_{fA\sigma} \rangle$ and $\langle n_{fB\bar{\sigma}} \rangle$ as well as the anomalous correlators possess just slight quantitative changes in comparison with the $U=0$ case. Thus, the effect of the Coulomb interaction is negligible in the case of $U<U_{c}$.

On the contrary, it follows from Figs. \ref{2}c,d that at $U>U_{c}$ the $C_{4}$ symmetry becomes spontaneously broken. Along with that the occupation of the sites becomes unequal for the different spin projections. The plots emphasize the essential role of the corners in this effect. It can be concluded with good accuracy that the average concentration deviates from unity only at these sites. Because of the Coulomb repulsion, the two distributions, $\langle n_{fA\uparrow} \rangle + \langle n_{fB\downarrow} \rangle$ and $\langle n_{fA\downarrow} \rangle + \langle n_{fB\uparrow} \rangle$, are the mirror images of each other. Interestingly, the anomalous correlators acquire an imaginary component which makes the main contribution again in the corners.

The crossover appears due to the competition between the Coulomb repulsion contribution to the ground-state energy and the contribution due to the overlapping of the excitations localized in the different corners. Thus, the $U_c$ value is dependent on the system size (for $N=20$ the curve break in Fig. \ref{1} emerges already at $U_{c}\approx0.04$) and becomes zero at the $N\rightarrow\infty$ limit.

The obtained results were proved by means of the ground-state energy analysis,
\begin{eqnarray}
	\label{Egr_weak}
&&E_{gr}=-\sum_{f\eta n\sigma}|v_{f\eta n\sigma}|^{2}\varepsilon_n-\\
&&~~~~~~~~~~~~~~~-U\sum_{f\eta\sigma}\left[\langle n_{f\eta\uparrow}\rangle\langle n_{f\eta\downarrow}\rangle +|\langle c_{f\eta\uparrow}^{+}c_{f\eta\downarrow}^{+}\rangle|^2\right].\nonumber\,
\end{eqnarray}
It was done for the fully-symmetric case, when the normal correlators are spin-independent and coincide in all corners, and for a set of the spin-asymmetric realizations. The last includes the situations when the same-spin normal correlators are equal in the two opposite corners of the square diagonal, in the two corners on the same square side, in the three and four corners. The minimum energy corresponds to the fully-symmetric solution for $U<U_c$ and the $C_2$-symmetric case with the same correlators in the opposite corners of the square diagonal for $U>U_c$.

\section{Strong correlation regime\label{sec4}}

\subsection{Effective low-energy interactions\label{sec4.1}}

Having discussed the limit of the weak Coulomb interaction, let us consider the properties of corner modes in the strong correlation regime. In this case, the Hartree-Fock approximation (\ref{HU_weak}) becomes invalid and it is necessary to use the methods of the theory of strongly correlated systems. First of all, we note that strong electron correlations induce effective interactions in low-energy Hamiltonian. Recently the effective interactions have been studied in interacting topological insulators \cite{rachel-18} and first-order topological superconductors \cite{zlotnikov-20}. To analyze the structure of effective interactions of the system (\ref{H01}), it is convenient to use the method of unitary transformations in many-body Hilbert space \cite{bir-72} together with the atomic representation \cite{hubbard-65, ovchinnikov-04}. This approach is described in Appendix \ref{apxB}. Since the natural language of the atomic representation is based on the use of Hubbard operators, $X^{pq}_{f\eta}$, we introduce two-component field operators, Hubbard spinors, built on such operators,
\begin{eqnarray}
\label{psi}
\Psi_{f\eta}=\left( \begin{array}{*{20}{c}}
X^{0\uparrow}_{f\eta}  \\
X^{0\downarrow}_{f\eta}
\end{array} \right)=
P\left(\begin{array}{*{20}{c}}
c_{f\eta\uparrow}\\
c_{f\eta\downarrow}
\end{array}\right)P-P\left(\begin{array}{*{20}{c}}
n_{f\eta\downarrow}\\
n_{f\eta\uparrow}
\end{array}\right)P,
\end{eqnarray}
where the Hubbard operators, $X^{pq}_{f\eta}$, and the projection operator, $P$, are defined in (\ref{P}). We will associate the operators  $X^{0\sigma}_{f\eta}$ constituting these spinors with so-called Hubbard fermions. It can be seen from Eq.\,(\ref{psi}) that in actual Hilbert space the Hubbard fermions are a superposition of the ordinary fermions, $c_{f\eta\sigma}$, and charge population operators $n_{f\eta\bar{\sigma}}$. As a result, the commutation relations for the Hubbard fermions differ from the ones for the ordinary fermions, which is the reason for the appearance of the kinematic interaction \cite{dyson-56, ivanov-88}. Another consequence of unusual operator algebra is the emergence of effective charge and magnetic interactions for itinerant electrons. So, using the Eq.(\ref{X2S}) it can be checked that
\begin{equation}\label{tau2S}
\Psi^+_{f\eta}\,\Psi_{f\nu} = \delta_{\eta\nu}\,{n}_{f\eta}\,,~~\Psi^+_{f\eta}\,\vec{\tau}\,\Psi_{f\nu} = 2\delta_{\eta\nu}\,\vec{S}_{f\eta}.
\end{equation}
where $\vec{\tau}$ is a vector consisting of the Pauli matrices acting in the spin space of the Hubbard fermions, $n_{f\eta}$ and $\vec{S}_{f\eta}$ are the charge and spin operators defined at the site $f$ and orbital $\eta$, respectively.

In terms of the spinors (\ref{psi}), the low-energy Hamiltonian obtained in the second-order perturbation theory (with $1/U$ as an expansion parameter) can be represented in the form (\ref{Heff}).
If $f\neq g \neq l$ the terms in lines 3-8 of (\ref{Heff}) correspond to three-center interactions. Their physical meaning consists in the hopping and anomalous pairing of Hubbard fermions at the $f$th and $g$th sites with a contact interaction at the $l$th site. In the lines 3, 4, and 5 of (\ref{Heff}) such interactions have the Coulomb, Heisenberg, and Dzyaloshinskii-Moriya character, respectively. These couplings possess an amplitude $\sim 1/U$ and can be realized both between the same orbitals (which is denoted by a factor $\delta_{\eta\nu}$) and between different orbitals (see a factor $\delta_{\bar{\eta}\nu}$). In the line 6 of (\ref{Heff}) the three-center interaction has an order $\sim \alpha^2/U$ and is related to the anisotropic hopping of Hubbard fermions. Similarly, the effective interactions with magnitudes $\sim \alpha\, \Delta_1\,/\,U$, written in the lines 7-8, describe the anisotropic interaction of Cooper pairs of the Hubbard fermions with the spin moments of the electrons at the site $l$. The anisotropy is due to the chirality of the spin-orbit interaction in Eq.\,(\ref{H01}).

\begin{multline} \label{Heff}
    \mathcal{H}_{eff}=P\,H\,P - \frac{1}{2}\,P\,\left(\bar{{\mathcal{V}}}\,\left(\mathcal{H}_{0} - K\mathcal{H}_{0}K\right)^{-1}\,\bar{{\mathcal{V}}} + h.c.\right)P\\
    =\sum_{f=1}^{N}\sum_{\eta=A,B}\sum_{\sigma=\uparrow, \downarrow}\left(-\mu + \eta \Delta \varepsilon\right)n_{f\eta\sigma} - \sum\limits_{\langle f\,g\,l \rangle}\sum_{\eta\,\nu}\\
    \Biggl\{\frac{1}{4U}\,\Bigl[\left(t_{fl}t_{lg}-\Delta_{fl}\Delta_{lg}\right)\delta_{\eta\nu}+\alpha^2\delta_{\bar{\eta}\nu}\Bigr]\,\Psi^+_{f\eta}\,n_{l\nu}\,\Psi_{g\nu}-\\
    -\frac{1}{U}\,\Bigl[\left(t_{fl}t_{lg}+\Delta_{fl}\Delta_{lg}\right)\delta_{\eta\nu}-\alpha^2\delta_{\bar{\eta}\nu}\Bigr]\,\Psi^+_{f\eta}\,\vec{\tau}\cdot \vec{S}_{l\nu}\,\Psi_{g\nu}-\\
    -\frac{2\alpha\,t_{gl}}{U}\,\delta_{\bar{\eta}\nu}\left(\vec{e}_z\times \vec{e}_{fg}\right)\,\Psi^+_{f\eta}\,\vec{\tau}\times \vec{S}_{l\eta}\,\Psi_{g\nu}+\\
    +\frac{2\alpha^2}{U}\delta_{\bar{\eta}\nu}\left(\vec{S}_{l\bar{\eta}}\times \vec{e}_{fg}\right)_z\Psi^+_{f\eta}\,\left(\vec{\tau}\times \vec{e}_{fg}\right)_z\,\Psi_{g\nu}-\\
    - \frac{2\alpha\,\Delta_{gl}}{U}\,\delta_{\bar{\eta}\nu}\,\Psi^+_{f\eta}\big[\,e^x_{fl}\,\big(\tau_x\,S^x_{l}-\tau_y\,S^y_l+\tau_z\,S_l^z\big)-\\
    -e^y_{fl}\,\big(\,\tau_x\,S_l^y+\tau_y\,S_l^x\,\big)\,\big]\Psi^+_{g\nu} + h.c.\, \Biggr\} + \\
    \sum_{\langle fg \rangle \eta} \Biggl\{ \left(t_{fg}\eta +\frac{\Delta_{fg}\,\Delta_0}{U}\right)\Psi^+_{f\eta}\,\Psi_{g\eta} + i\alpha\,\Psi^+_{f\eta}\,\vec{\tau}\times \vec{e}_{fg}\,\Psi_{g\bar{\eta}}\\
    +\Psi^+_{f\eta}\left(i\Delta_{fg}\tau_y-\frac{t_{fg}\Delta_0\eta}{U}\tau_x\right)\Psi^+_{g\eta}-\frac{i\alpha\Delta_0}{U}\,e_y\,\Psi^+_{f\eta}\,\tau_z\,\Psi^+_{g\bar{\eta}} -\\
    - \frac{\Delta^2_{fg}}{U}\left(\vec{S}_{f\eta}\vec{S}_{g\eta}-\frac{1}{4}n_{f\eta}n_{g\eta}\right)+h.c.\Biggr\}.
\end{multline}

It is important to note, that if $f=g$ the three-center terms reduce to the two-center charge and spin interactions between the electrons, according to Eq.\,(\ref{tau2S}). So, the two-center summands in the third line of Eq.(\ref{Heff}) describe the intersite Coulomb repulsion $\sim n_{f\eta}n_{g\nu}$ of the electrons inside the same orbitals which is formed by the competition of attractive and repulsive interactions with amplitudes $\sim t^2/U$ and $\sim \Delta^2/U$, respectively.
Similarly, the symmetric Heisenberg interaction $\sim \vec{S}_{f\eta}\cdot \vec{S}_{g\nu}$ is realized inside the orbitals and has an antiferromagnetic character with an amplitude $\sim t^2/U$. Note that the Dzyaloshinskii-Moriya terms $\sim \vec{S}_{f\eta}\times \vec{S}_{g\nu}$ as well as the anisotropic two-center interactions do not appear, since the spin-orbit interaction acts only between the different orbitals in the original model (\ref{H01}). The discussed interactions can lead to the implementation of charge and spin orderings, which, in turn, should be taken into account when calculating the matrix elements of the three-center interactions. Thus, the Hubbard fermions move in the charge and magnetic background.

The two-center terms given in the second curly brackets of Eq.(\ref{Heff}) describe the hopping, spin-orbit interaction and anomalous pairings between the nearest neighbors in the ensemble of the Hubbard fermions. It can be seen that the interorbital spin-orbit interaction induces a p-wave superconducting pairing between the neighboring orbitals, similar to what occurs in Majorana nanowires.

The results presented show that the search for the Majorana corner modes in the regime of strong but finite $U$ requires to study of spectral properties of the system taking into account the magnetic ordering, intersite repulsion, p-wave anomalous pairing, anisotropic hoppings as well as the three-center and kinematic interactions. Such an analysis is beyond the scope of this work. Meanwhile, it is clear that in the limit $U \to \infty$ one can consider only the influence of the kinematic interaction on the MCM implementation conditions.

\subsection{\label{sec4.2} $U \to \infty$ limit}

In the $U \to \infty$ limit the system is described by two bands corresponding to the lower Hubbard subbands for the Ath and Bth orbitals. Here we consider the case when the bare energies of the Ath and Bth orbitals are shifted by the parameter $\Delta \varepsilon \ne 0$, while the intraorbital hopping between the next-nearest neighbors with the parameter $t_1$ is neglected for simplicity. Then, in the $U \to \infty$ limit the Hamiltonian \eqref{Heff} can be written as
\begin{eqnarray}
\label{Hinfty}
&& {\mathcal{H}}_{U \to \infty} = PHP =
\sum_{f \sigma} \sum_{\eta = A, B} \left( -\mu + \eta \Delta \varepsilon \right) X_{f \eta}^{\sigma \sigma}
\nonumber \\
& + & \sum_{f \eta \sigma} \sum_{\delta = \pm x, \pm y} \eta t_{\delta} X_{f \eta}^{\sigma 0} X_{f+\delta, \eta}^{0 \sigma} + \sum_{f \delta \eta \sigma} \alpha_{\sigma \delta} X_{f \eta}^{\sigma 0} X_{f+\delta, \bar{\eta}}^{0 \bar{\sigma}}
\nonumber \\
& + & \sum_{f \delta \eta} \left( \Delta_1 X_{f \eta}^{\uparrow 0} X_{f+\delta, \eta}^{\downarrow 0} + h.c. \right),
\end{eqnarray}
where the orbital index $\bar{\eta} = B(A)$ if $\eta = A(B)$, respectively. As before $t_{\pm x} = -t_{\pm y} = t$, $\alpha_{\sigma, \pm x} = \mp \alpha \sigma$, $\alpha_{\sigma, \pm y} = \pm i\alpha$.

Obviously, in the $U \to \infty$ limit the on-site singlet pairing is fully suppressed by the local Coulomb repulsion. Therefore, the parameter $\Delta_0$ does not appear in the Hamiltonian \eqref{Hinfty} and the topological phase transition to the trivial phase shown in Fig. \ref{proc_fig} becomes inaccessible.

\begin{figure*}[tb]
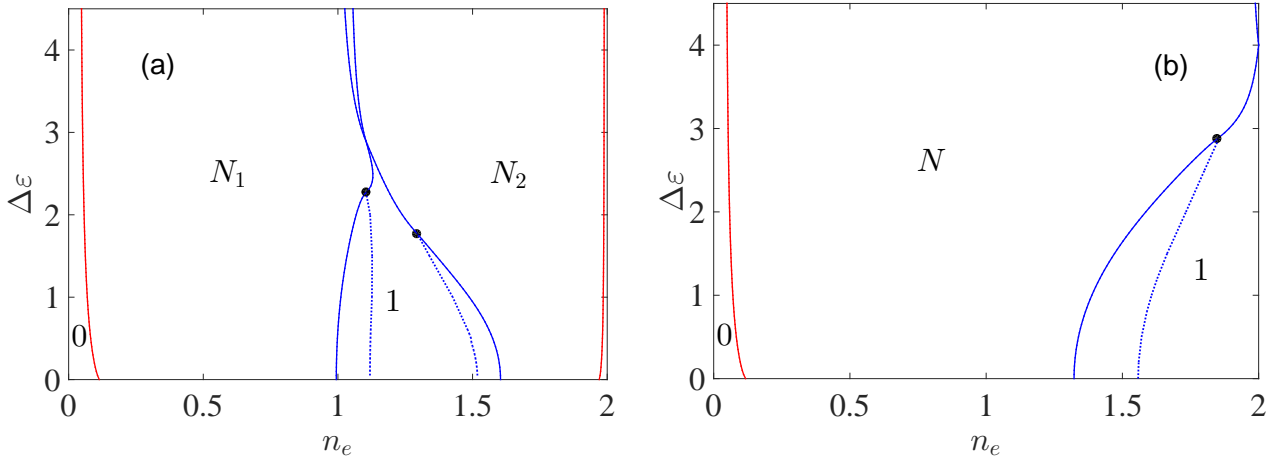

    \includegraphics[width=0.45\textwidth]{fig4a.eps} \, \, \,
    \includegraphics[width=0.45\textwidth]{fig4b.eps}
    \caption{\label{fig_tp} (a) The topological phase diagram in the $U \to \infty$ limit in the variables $\Delta \varepsilon$ and electron concentration $n_e = n_A + n_B$. $N_1$ and $N_2$ denote the nodal phases in which the bulk excitation spectrum is gapless and edge or corner modes are prohibited. The maximum concentration in this limit is $n_e = 2$. The phase with the notation $0$ is a gapped topologically trivial phase. The same phase is found in the vicinity of $n_e = 2$. The notation $1$ marks the topologically nontrivial phase where the Majorana corner modes are realized. The dotted lines are the conditions when the edge excitation spectra (along (10) or (01) edges) are gapless. (b) The topological phase diagram for $U = 0$. This phase diagram can be symmetrically continued to the $n_e = [2-4]$ range. The parameters are $t=1$, $\alpha = 3/4$, $\Delta_1 = 0.5$, $t_1 = 0$}
\end{figure*}

Using the formalism of the Zubarev's Green functions (see Appendix \ref{apxC} ) the topological phase diagram is considered in the limit of $U \to \infty$ within the Hubbard-I approximation.
Firstly, we are focused on the boundaries of nodal phases ($N$ phases) in which the gapless excitations exist in the bulk spectrum due to the $s_{\pm}$ symmetry of the superconducting pairings. To find the $N$ phases the periodic boundary conditions have to be applied with the uniform correlators determining Hubbard renormalizations.

In general, the gapless excitations appear when the Fermi contour intersects the nodal lines of the superconducting order parameter. Since the on-site superconducting pairings are suppressed in the limit of $U \to \infty$, the nodal lines are determined by simple relations: $k_{cy} = \pm(\pi - |k_{cx}|)$. Therefore, to describe the nodal phases we found the conditions when the zeros on the nodal lines in the bulk energy spectrum of topological insulator (TI) appear.

The bottom of the first TI band $\varepsilon_{1k}$ and the top of the second TI band $\varepsilon_{2k}$ (see Appendix \ref{apxC}) are realized at the nodal points $k_{cx} = 0$, $k_{cy} = \pm \pi$. Then, the condition
\begin{equation}
\label{muL1}
\mu_{L1} =  -\Delta \varepsilon - 4tH_{B}
\end{equation}
is the lower boundary of the nodal phase corresponding to the filling of $\varepsilon_{1k}$ (the $N_1$ phase), while
\begin{equation}
\label{muU2}
\mu_{U2} =  \Delta \varepsilon + 4tH_{A}
\end{equation}
is the upper boundary of the nodal phase corresponding to the filling of $\varepsilon_{2k}$ (the $N_2$ phase).
Here $H_{\eta} = 1 - n_{\eta}/2$ is the Hubbard renormalization, $n_{\eta} = \sum_{\sigma} \left\langle X_{f \eta}^{\sigma \sigma} \right \rangle$ is the average electron concentration at the $\eta$th orbital (it does not depend on the site index since the periodic boundary conditions are considered), $\eta = A, B$. The concentrations of the Hubbard fermions with the different spins are equal. We note that in the limit of $U \to \infty$ the electron concentration on each orbital can not exceed 1.

\begin{figure}[tb]
    \includegraphics[width=0.3\textwidth]{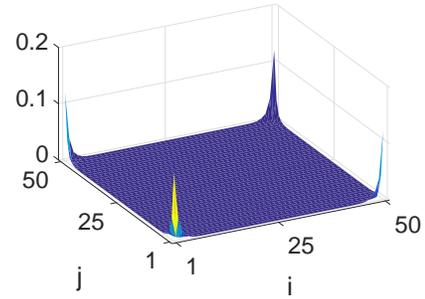}
    \caption{\label{MCM_Uinf} Probability density of the Majorana corner modes in the topologically nontrivial $1$ phase in the $U \to \infty$ limit on the 2D lattice with $N = 50$.}
\end{figure}

\begin{figure*}[tb]
    \includegraphics[width=0.45\textwidth]{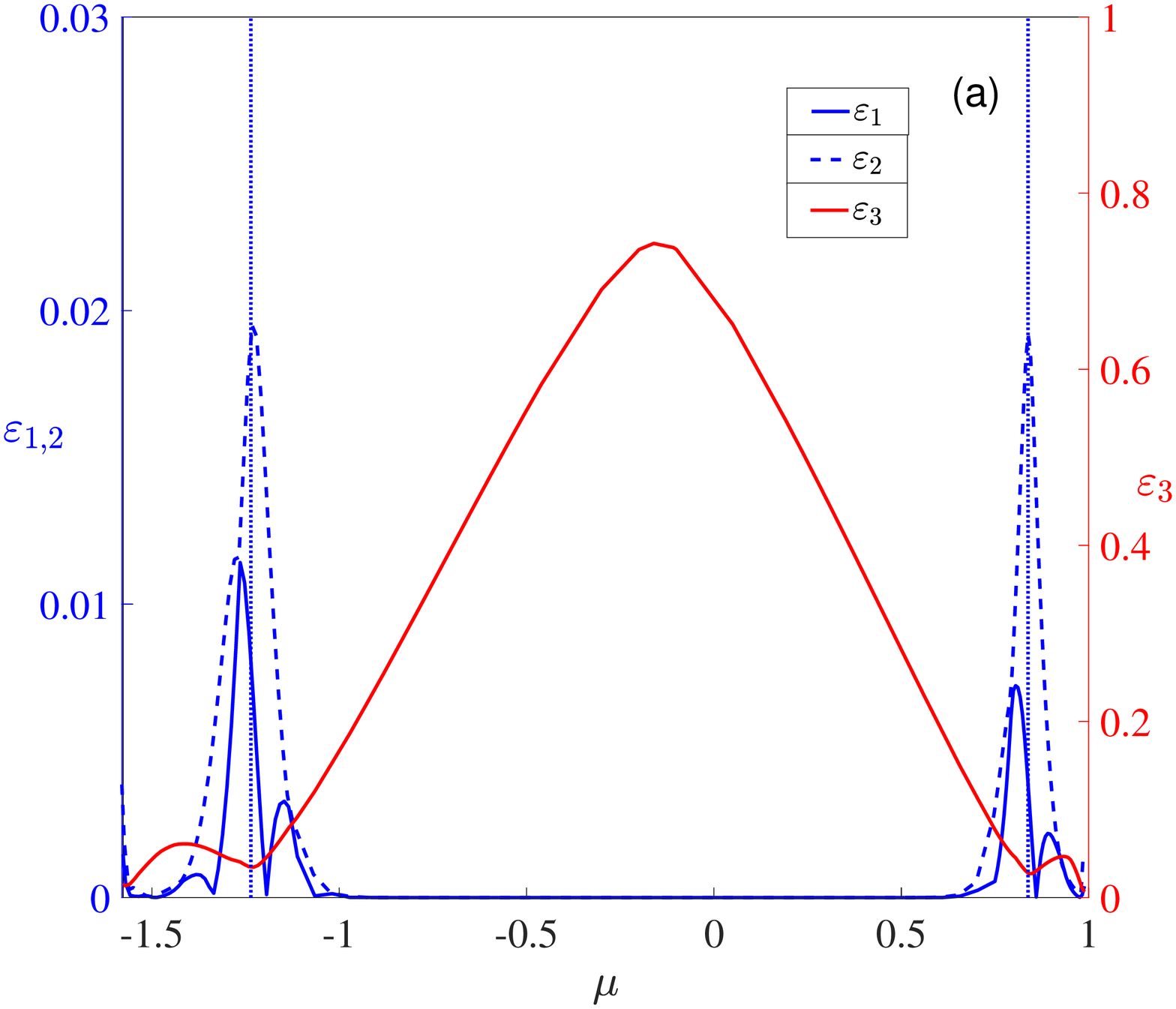} \, \, \,
    \includegraphics[width=0.45\textwidth]{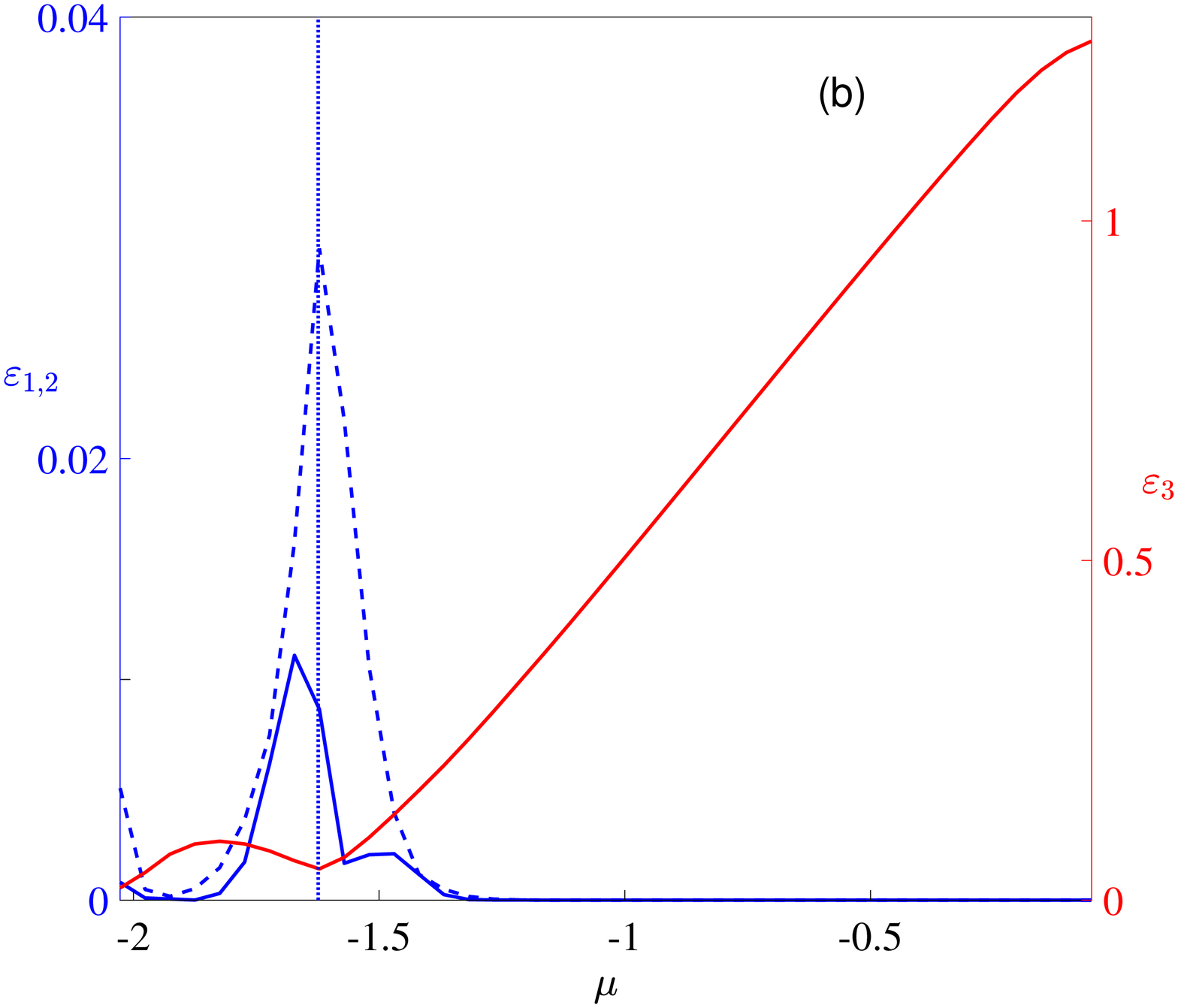}
    \caption{\label{fig_ejmu} (a) The dependencies of the three lowest excitation energies on the chemical potential inside the $1$ phase from Fig. \ref{fig_tp}a at $\Delta \varepsilon = 1$ and $U \to \infty$. Left y-axis is for the energies $\varepsilon_{1,2}$ which become zero on the interval $\mu \in \left[-1,0.7\right]$. Right y-axis is for $\varepsilon_{3}$ which determines the energy gap for the zero modes. As in Fig. \ref{fig_tp}, the edge excitation spectra are gapless for the chemical potentials denoted by the vertical dotted lines. (b) The case of $U = 0$.}
\end{figure*}

The upper boundary of the
$N_1$ phase and the lower boundary of the $N_2$ phase are described by the expressions
    \begin{eqnarray}
\label{U1L2}
     \mu_{U1, L2} & = & -\frac{t^2 \Delta \varepsilon \left( H_{A}^2-H_{B}^2 \right)}{2\left[ t^2 \left( H_{A}+H_{B} \right)^2/2 - H_{A}H_{B}\alpha^2 \right]} \pm
    \nonumber \\
    &\mp& \left\{  \frac{8H_{A}^2H_{B}^2\alpha^2(2t^2-\alpha^2) - H_{A}H_{B} \Delta \varepsilon^2 \alpha^2}{t^2 \left( H_{A}+H_{B} \right)^2/2 - H_{A}H_{B}\alpha^2} \right.
    \nonumber \\
    &+& \left.\frac{H_{A}H_{B}\left( H_{A}-H_{B} \right)^2 \Delta\varepsilon^2 \alpha^2 t^2}{2\left[t^2 \left( H_{A}+H_{B} \right)^2/2 - H_{A}H_{B}\alpha^2 \right]}\right\}^{1/2},
    \end{eqnarray}
when the Fermi contour intersects the points on the nodal lines determined by
    \begin{eqnarray}
    \label{ineq_par}
    \cos(k_{cx}) & = & - \frac{t \left[ \mu \left( H_{A}-H_{B} \right) + \Delta \varepsilon \left( H_{A}+H_{B} \right) \right]}{4 H_{A}H_{B}\left( 2t^2 - \alpha^2 \right)} > -1,
    \nonumber \\
    k_{cy} & = & \pm(\pi - |k_{cx}|).
    \end{eqnarray}
When $\cos(k_{cx}) < -1$ upon changing the parameters, the upper boundary of the $N_1$ phase corresponding to the top of $\varepsilon_{1k}$ and the lower boundary of the $N_2$ phase corresponding to the bottom of $\varepsilon_{2k}$  are implemented at the points $k_{cx} = \pm \pi$, $k_{cy} = 0$. Then, the conditions for the chemical potential read
\begin{eqnarray}
\label{muU1}
\mu_{U1} =  \Delta \varepsilon - 4tH_{A},
\\
\label{muL2}
\mu_{L2} =  -\Delta \varepsilon + 4tH_{B}.
\end{eqnarray}

In the obtained expressions $(\ref{muL1}-\ref{muL2})$ for the boundaries of the nodal phases the average electron concentrations $n_{\eta}$ included in the renormalization parameters $H_{\eta}$ must be calculated self-consistently. The self-consistent equations and expressions for the bulk energy spectrum of HOTSC are provided in Appendix \ref{apxC}. Note that the boundaries of the $N$ phases in $U = 0$ case can be found from these expressions neglecting the Hubbard renormalizations, $H_{A} = H_{B} = 1$.

In Figures \ref{fig_tp}a and \ref{fig_tp}b we present the topological phase diagrams in the variables $\Delta \varepsilon$ and electron concentration $n_e = n_A + n_B$ for the limit of $U \to \infty$ and for the $U = 0$ case, respectively. The parameters $t=1$, $\alpha = 3/4$, $\Delta_1 = 0.5$, $t_1 = 0$ are used. For clarity, we put $\Delta_0 = 0$ in the $U = 0$ case. The red solid lines are determined by Eqs. \eqref{muL1} and \eqref{muU2} for the $N_1$ and $N_2$ nodal phases, respectively. The blue lines are determined by Eqs. \eqref{U1L2} and (\ref{muU1}-\ref{muL2}) depending on the parameter range. The dots on these lines denote when $\cos(k_{cx})=-1$ in \eqref{ineq_par} and the equations for the phase boundaries are changed from \eqref{U1L2} to (\ref{muU1}-\ref{muL2}) with the increase of $\Delta \varepsilon$.

The notations for the different phases on the topological phase diagrams are the same as in Ref. \cite{Wang-18}.
As mentioned above, inside the $N_1$ and $N_2$ phases the bulk energy spectrum is gapless in the presence of the superconducting pairings and there are not edge or corner states.
The phases with the gapped bulk energy spectrum are $0$ and $1$ phases distinguished by topology. The topologically protected edge and corners states are absent in the topologically trivial $0$ phase. The Majorana corner modes are formed in the topologically nontrivial $1$ phase. In this phase the edge excitation spectra along (10) or (01) edges are gapped excepting the parameters shown by the dotted lines. As it was shown in Ref. \cite{Wang-18} for $U = 0$ and $\Delta_0 = 0$ the topological phase transition does not occur at this line. In the $U \to \infty$ limit we have the same result, since the on-site pairings are destroyed by the Coulomb interaction.

To compare the limits of $U \to \infty$ and $U=0$ in Fig. \ref{fig_tp}b the half of the topological phase diagram at $U=0$ is shown. The whole phase diagram is determined on the range  $n_e = [0-4]$ and it is symmetric relative to $n_e = 2$. It is seen from Fig. \ref{fig_tp} that all phases preserve in the $U \to \infty$ limit within the Hubbard-I approximation. At the same time, the phases are shifted to the lower concentrations and are compressed due to the Hubbard renormalizations. In Sections \ref{sec3.2} and \ref{sec3.3} the doping level near the half-filling $n_e = 2$ at $U=0$ is considered. It is seen in Fig. \ref{fig_tp}a that this region becomes topologically trivial if $U \to \infty$.

To check the topologically nontrivial $1$ phase the excitation spectrum $\varepsilon_j$ of the 2D lattice with open boundary conditions and the MCMs spatial distribution are calculated using the Green functions. The difference of the Hubbard renormalization factors at the different lattice sites is neglected and the bulk uniform values for them are used. The MCM formation deeply inside the $1$ phase is displayed in Fig. \ref{MCM_Uinf}. The lattice contains $N = 50$ sites along the $x$ and $y$ directions.
In Fig. \ref{fig_ejmu}a the dependencies of the lowest excitation energies $\varepsilon_{j}$ on the chemical potential at $\Delta \varepsilon = 1$ in the $1$ phase are presented. The chemical potential runs from the left boundary of the $1$ phase to the right boundary. The other parameters remain the same. Since the scales of the energies $\varepsilon_{1,2}$ and $\varepsilon_{3}$ are different, we employ the different y-axes for them (the left y-axis is for $\varepsilon_{1,2}$, the right y-axis is for $\varepsilon_{3}$). For the chemical potentials denoted by the vertical dotted lines the edge excitation spectrum is gapless. It is seen that two zero excitation energies $\varepsilon_{1,2}$ corresponding to the MCM formation are realized in a wide range of the chemical potential excepting the regions near the vertical dotted lines. In Fig. \ref{fig_ejmu}b the results for $U=0$ are shown. Comparing the dependencies of $\varepsilon_3$ in both cases, we conclude that the energy gap between the MCMs and higher states is slightly decreased in the $U \to \infty$ regime.

\section{\label{sec5}Summary}

The effect of the on-site Coulomb interaction on the HOTSC was investigated on the example of the topological insulator with enhanced $s(d)$-wave superconducting coupling in two regimes: weak and strong Coulomb repulsion. Using the mean-field approximation in the weak regime it was shown that the on-site intraorbital Coulomb interaction manifests itself only in modification of the on-site energy shift and suppression of the on-site singlet superconducting coupling. In the uniform case it leads to the widening of the higher-order topological phase.

When the self-consistent solution takes into account the boundary of the finite size system the conventional topological analysis becomes invalid since, in this case, the correlators are site-dependent leading to the inhomogeneous picture.
Meanwhile the corner excitations survive in this case. The crossover between two different situations was found. If the amplitude of the Coulomb repulsion is less than the critical value, the corner excitation energies are determined by the hybridization effects due to the finite size of the system. The electron densities for different spin projections are equal and $C_4$-symmetric in this case. If the Coulomb repulsion is stronger then the critical value, the spontaneous symmetry breaking emerges in the system and the corner excitation energy depends quadratically on $U$. The electron densities for different spin projections are $C_2$ symmetric with the difference taking place in the corners of the system. This crossover is a finite-size effect appearing at the lesser $U$ for the larger system size $N$.

The effective interactions in the strongly correlated HOTSC are derived in the framework of the second-order operator-form perturbation theory. The appearance of antiferromagnetic and ferromagnetic exchange interactions, anisotropic interactions, as well as triplet pairings are demonstrated. It is shown that the topologically nontrivial phase in the vicinity of on-site electron concentration $n_e = 2$ (half-filling case at $U = 0$) becomes trivial one in the strongly correlated regime. On the other hand, in this regime the lower Hubbard subbands for both orbitals behave qualitatively similar to the initial bands without the Coulomb interaction. Therefore, the topological phase can be realized even at $U \to \infty$. At the same time, the topological region on the phase diagram in variables electron concentration --- orbital splitting, as well as the energy gap for the corner states are reduced due to the Hubbard renormalizations.

\begin{acknowledgments}
We acknowledge fruitful discussions with D.M. Dzebisashvili and V.A. Mitskan. The reported study was supported by Russian Science Foundation, project No. 22-22-20076, and Krasnoyarsk Regional Fund of Science.
\end{acknowledgments}

\appendix

\section{\label{apxA} HOTSC phase diagram employing effective mass criterion}

To analyze the conditions of the HOTI/HOTSC phase realization it is useful to employ an effective mass criterion. It can be introduced if the system possesses topological edge states, which are gapped under the influence of some perturbations \cite{Langbehn-17,Zhu-18,Wang-18,Yan-18,zhang-20PRR,zhang-20PRB,Wu-20,Ikegaya-21,Fedoseev-22}. In such case the HOTSC phase appears when the effective Dirac mass of the edge excitations is of a different sign for two adjacent edges. To use the effective mass sign criterion in our case one needs to find the edge eigenstates of the Hamiltonian (\ref{H01}) in the absence of the superconducting coupling with one open boundary. Let us consider the boundary along the $x$ direction. The edge-state wave function in such case can be written in the form
\begin{eqnarray}
\label{WF_el}
&&\Psi_{ep\sigma}=\frac{1}{\sqrt{2\mathcal{N}}}\left[\begin{array}{c} 1 \\ -i\cdot \sign(\alpha t_p)\end{array}\right]\cdot\left(x_1^n-x_2^n\right),\\
&&x_{1,2}=-\frac{\xi_p\pm\sqrt{\xi_p^2-4(t_p^2-\alpha^2)}}{2\sign(t_p)(|t_p|+|\alpha|)},\nonumber\\
&&\xi_p=\Delta\varepsilon+2t_x\cos p,~~t_p=t_y+2t_1\cos p,\nonumber\\
&&\mathcal{N}=\sum_{n=1}^{\infty}\left|x_1^n-x_2^n\right|^2,\nonumber
\end{eqnarray}
with edge band energy spectrum
\begin{eqnarray}
\label{Ener_el}
&&\varepsilon_p=2|\alpha|\sign(t_p)\cdot\sigma\sin p.
\end{eqnarray}
Here the basis $[c_{pA\sigma},c_{pB\overline{\sigma}}]^T$ is used, $p=k_x$ is quasi-momentum along the boundary, an index $n$ numerates the sites in $y$ direction. The values $x_{1,2}$ can be both real or complex (in the last case $x_2=x_1^*$), along with $|x_{1,2}|<1$ corresponding to the solution, which descends along $y$ direction inside the system.

The hole-like counterpart of (\ref{WF_el}) in the $[c_{pA\overline{\sigma}}^{\dag},c_{pB\sigma}^{\dag}]^T$ basis has the form
\begin{eqnarray}
\label{WF_hl}
&&\Psi_{hp\sigma}=\frac{1}{\sqrt{2\mathcal{N}}}\left[\begin{array}{c} 1 \\ i\cdot \sign(\alpha t_p)\end{array}\right]\cdot\left(x_1^n-x_2^n\right),\\
&&\varepsilon_p=-\sign(\alpha t_p)\cdot\sigma\alpha_p.\nonumber
\end{eqnarray}

Referring to (\ref{WF_el}),(\ref{WF_hl}) as electron and hole wave-functions and projecting the whole Hamiltonian (\ref{H01}) on these lowest-energy solutions, one will obtain the next form
\begin{eqnarray}
\label{Ham_proj}
&&\mathcal{H}_{pr}=\left[\begin{array}{cc} \varepsilon_p-\mu & V_p^* \\ V_p & -\varepsilon_p+\mu\end{array}\right],\\
&&V_p=\sigma\left(\Delta_p-2\Delta_y\frac{\xi_p \sign(t_p)}{2\sqrt{t_p^2-\alpha^2}}\right),\nonumber\\
&&\Delta_p=\Delta_0+2\Delta_x\cos p.\nonumber
\end{eqnarray}

The excitation spectrum of Hamiltonian (\ref{Ham_proj}) is Dirac-like,
\begin{eqnarray}
\label{E_proj}
&&\varepsilon=\sqrt{(\varepsilon_p-\mu)^2+|V_p|^2},
\end{eqnarray}
around the Dirac point defined by equation
\begin{eqnarray}
\label{E_proj}
&&\sign(t_p)\sin p=\mu/2\alpha,
\end{eqnarray}
and $V_p$ playing a role of the effective Dirac mass.

The wave functions of the edge states on the $y$ boundary with $p=-k_y$ has form
\begin{eqnarray}
\label{WF_y}
&&\Psi_{ep\sigma}=\frac{1}{\sqrt{2\mathcal{N}}}\left[\begin{array}{c} 1 \\ \sigma \sign(\alpha t_p)\end{array}\right]\cdot\left(x_1^n-x_2^n\right),\\
&&\Psi_{hp\sigma}=\frac{1}{\sqrt{2\mathcal{N}}}\left[\begin{array}{c} 1 \\ -\sigma \sign(\alpha t_p)\end{array}\right]\cdot\left(x_1^n-x_2^n\right),\nonumber
\end{eqnarray}
with all other expressions including $V_p$ to be the same up to the $x\leftrightarrow y$ exchange. The HOTSC phase appears in the case of $V_p$ having different signs for $x$ and $y$ boundaries at the corresponding Dirac points. Supposing the system with hopping amplitudes and intersite superconducting coupling to differ only in signs $t_x=\kappa t_y$, $\Delta_x=\chi\Delta_y$ ($\kappa,\chi=\pm1$) one will easily find the requirement $\kappa\chi=-1$ for the HOTSC phase, which coincides with the conclusions made in \cite{Wang-18} ($t_x=-t_y$ for the extended $s$-wave superconducting coupling and $d_{x^2-y^2}$-wave for $t_x=t_y$).

In the case of $\Delta\varepsilon<|t_{x,y}|$, $|t_{x,y}|=2|t_1|$ (the situation considered in \cite{Wang-18}), the HOTSC phase is defined by the condition
\begin{eqnarray}
\label{HOTSC_cond1}
&& \left|\frac{\Delta_0}{2\Delta_1}\right|<\left|
\left(1+\frac{1}{\sqrt{\left(1+|\cos p|\right)^2-\left(\frac{\alpha}{t_x}\right)^2}}\right)\cdot|\cos p|\right.\\
&&+\left.\frac{\Delta\varepsilon\cdot \sign(t_xt_y)}{2|t_x|\sqrt{\left(1+|\cos p|\right)^2-\left(\frac{\alpha}{t_x}\right)^2}}\right|, ~~|\cos p|=\sqrt{1-\left(\frac{\mu}{2\alpha}\right)^2}.\nonumber
\end{eqnarray}
The obtained expression can describe HOTSC phase only in the case of $|\mu|<2\alpha$ as it is based on the perturbed edge states conception and, consequently, the chemical potential should be inside the edge states band (\ref{Ener_el}).

\section{\label{apxB} Effective Interactions in Strongly Correlated Regime}

Let us rewrite the original Hamiltonian, as a sum of terms of zero and first order of smallness:
\begin{eqnarray}\label{H_tot}
H = \mathcal{H}_{0} + \mathcal{V}.
\end{eqnarray}
Here $\mathcal{H}_{0}$ is an unperturbed Hamiltonian and $\mathcal{V}$ is an operator corresponding to the weak interactions. These operators can be represented in the form
\begin{multline}\label{H0_V}
\mathcal{H}_0 = \sum_{f\,\eta\,\sigma}\left(\,-\mu + \eta\,\Delta\varepsilon\right)c^+_{f\eta\sigma}c_{f\eta\sigma} + U\sum_{f\eta}n_{f\eta\uparrow}n_{f\eta\downarrow}\,,\\
\mathcal{V} =\\ \sum_{fg\,\eta\sigma\sigma'}\Big(t_{fg}\eta\,\delta_{\sigma\sigma'}\, c^{+}_{f\eta\sigma}c_{g\eta\sigma'}\nonumber
+i\alpha\,\left(\tau_{\sigma\sigma'}\times e_{fg}\right)_{z}\,c^{+}_{f\eta\sigma}c_{g\bar{\eta}\sigma'}+\\
+\frac{1}{2}\Delta_{fg}\,\sigma\,c^{+}_{f\eta\sigma}c^{+}_{g\eta\bar{\sigma}}\Big)+\Delta_{0}\sum_{f\eta}c^{+}_{f\eta\uparrow}c^{+}_{f\eta\downarrow}+h.c.
\end{multline}
Note that here we consider a general case in which the hopping and SC pairings can take place for distant neighbors with amplitudes $t_{fg}$ and $\Delta_{fg}$, respectively.

As a basis in the Hilbert space of the operator $H$ it is convenient to choose many-body eigenstates $|\,m\,\rangle$ of the Hamiltonian $\mathcal{H}_{0}$: $\mathcal{H}_{0}|\,m\,\rangle=E_m|\,m\,\rangle$.
An important assumption for the development of the perturbation theory is the existence of a large energy gap in the spectrum of the eigenvalues $E_m$. If we consider the system in the regime of the strong electron correlations, $$U\gg t_{fg},\,\Delta_{fg},\,\alpha,$$
the energy gap occurs due to the presence of the strong Hubbard repulsion. Then, the subspace of the states, $\mathcal{M}$, with the eigenvalues below the gap (so-called "low-energy"\ sector) include the ones without the doubly occupied orbitals at each site, i.e.
$$\mathcal{M} = \{  |\,m\,\rangle:~\forall f,\eta~~n_{f\eta}|\,m\,\rangle \neq 2\,|\,m\,\rangle  \};~n_{f\eta}=n_{f\eta\uparrow}+n_{f\eta\downarrow}.$$
The ''high-energy'' sector  $\mathcal{L}$ is formed by states $|l\rangle$ for which at least one orbital have two electrons.

Using the many-body states $|m\rangle$ we can define a projection operator $P$ onto the low-energy sector $\mathcal{M}$ as:
\begin{eqnarray}\label{P}
P = \prod_{f}\sum_{\eta=A,B}\left(X_{f\eta}^{00} + X_{f\eta}^{\uparrow\uparrow}+X_{f\eta}^{\downarrow\downarrow}\right),
\end{eqnarray}
with $X_{f\eta}^{pq}=|f\,\eta\,p\rangle\langle f\,\eta\,q|$ being the Hubbard operators describing transitions from the many-body state $|f\,\eta\,p\rangle$ to the state $|f\,\eta\,q\rangle$ at the site $f$ and orbital $\eta=A,B$, with quantum numbers $p$ and $q$, respectively. In our case the basis of states at the site $f$ and orbital $\eta$ includes $|f \,\eta\,0\rangle$, $|f\,\eta\,\sigma\rangle$ and $|f\,\eta\,2\rangle$ corresponding to the states without electrons, with one electron that has the spin $\sigma$ and with two electrons, respectively.
The electron annihilation operator at the site $f$ and orbital $\eta$ with spin projection $\sigma$ can be expressed in terms of the Hubbard operators:
\begin{equation}\label{Fermi2Hubbard}
c_{f\eta\sigma}=X_{f\eta}^{0\sigma}+\sigma X^{\bar{\sigma}2}_{f\eta}.
\end{equation}
The projection operator (\ref{P}) allows to divide the interactions into two parts: $\mathcal{V}=\mathcal{\bar{\bar{V}}}+\mathcal{{\bar{V}}}$, where
\begin{eqnarray}\label{H2}
\mathcal{\bar{{V}}} = (1-P)\,\mathcal{V}\,P+P\,\mathcal{V}\,(1-P),
\end{eqnarray}
is non-diagonal, since it does mix the sectors $\mathcal{M}$ and $\mathcal{L}$.

To derive the desired effective Hamiltonian of the strongly correlated HOTSC model, one can consider the following unitary transformation of the Hamiltonian $H$:
\begin{multline}\label{U_trans}
H\to \tilde{\mathcal{H}}=e^{-S}\,H\,e^{S}=H + [\,H, S\,] + \frac{1}{2}\Big[\,[\,H,\,S\,],\,S\,\Big] +\ldots
\end{multline}
It is assumed that the operator $S$ in the formula (\ref{U_trans}) is non-diagonal and has the first order of smallness. Next, it is necessary to substitute the expression (\ref{H_tot}) into the series (\ref{U_trans}) and retain only those terms whose order of smallness is not higher than two. In the obtained expression for $\tilde{\mathcal{H}}$ we want to get rid of the non-diagonal terms by imposing the following condition on the operators $S$:
\begin{eqnarray}\label{S1_eq}
\mathcal{\bar{{V}}} + [\, \mathcal{H}_{0}\, , \, S\,] = 0.
\end{eqnarray}

As a result, only the diagonal terms remain in the Hamiltonian $\tilde{\mathcal{H}}$ up to the second order.
Projecting out the high-energy processes in the last, we are left with operators acting exclusively within the low-energy sector $\mathcal{M}$ and, thus, forming the required effective Hamiltonian,
\begin{eqnarray}\label{H_eff}
\mathcal{H}_{eff} =P\,H\,P+ \frac{1}{2}\,P\, \big[ \,\mathcal{\bar{{V}}}\, , \, S\, \big]\,P + h.c.
\end{eqnarray}

It is easily to verify that $S$ can be represented in the operator form \begin{eqnarray}\label{S1_eq_op}
S = -\left(\mathcal{H}_{0} - K\mathcal{H}_{0}K\right)^{-1}\,(1-P)\,\mathcal{V}\,P - h.c.,
\end{eqnarray}
where $K$ is the Hermitian conjugation operator. Then, substituting the expression (\ref{S1_eq_op}) into the formula (\ref{H_eff}), we obtain the final expression for the effective Hamiltonian acting in the low-energy subspace $\mathcal{M}$:
\begin{multline}\label{H_eff_fin}
\mathcal{H}_{eff} = P\,\mathcal{H}\,P
-\frac{1}{2}\left(P\,\bar{
\mathcal{V}}\,\left(\mathcal{H}_{0} - K\mathcal{H}_{0}K\right)^{-1}\,\bar{\mathcal{V}}\,P + h.c.\right)
\end{multline}
In order to find the explicit microscopic expression for $\mathcal{H}_{eff}$ it is convenient to perform calculations representing $\mathcal{\bar{V}}$ in terms of Hubbard operators,
\begin{multline*}
    \bar{\mathcal{V}}
    =\sum_{\sigma\sigma'\eta\,fg}\\
    \Big( t_{fg}\delta_{\sigma\sigma'}\eta\sigma X^{\sigma 0}_{f\eta}X^{\bar{\sigma} 2}_{g\eta} + i\alpha\left(\vec{\tau}_{\sigma\sigma'}\times \vec{e}_{fg}\right)\sigma' X_{f\eta}^{\sigma 0}X_{g\bar{\eta}}^{\bar{\sigma}'2}+\\
    +\frac{\Delta_{fg}}{2}\,\delta_{\sigma\sigma'}\left(X_{f\eta}^{2\sigma}X_{g\eta}^{\sigma 0}- X_{f\eta}^{\sigma 0}X_{g\eta}^{2\sigma} + \sigma X_{f\eta}^{2{\sigma}}X_{g\eta}^{2\bar{\sigma}} \right) + h.c. \Big) + \\
    + \Delta_0\sum_{f\eta}\left(X_{f\eta}^{02} + X_{f\eta}^{20} \right),
\end{multline*}
and taking into account the relations \cite{ovchinnikov-04}
\begin{multline}\label{X2S}
X^{pq}_{f\eta}X^{rs}_{f\nu}=\delta_{\eta\nu}X_{f\eta}^{pr},\\X^{\sigma\sigma}_{f\eta}=n_{f\eta}+2\sigma\,S^z_{f\eta}\,,~ X_{f\eta}^{\sigma\bar{\sigma}}=2\left(S^x_{f\eta} + i\sigma\,S^y_{f\eta}\right).
\end{multline}

\section{\label{apxC} Green functions approach in the $U \to \infty$ limit}

The equation of motion for the operator $X_{f\eta}^{0 \sigma}(t)$ in the Heisenberg representation and for the Hamiltonian \eqref{Hinfty} is expressed in the Hubbard-I approximation as
\begin{eqnarray}
\label{EqM}
&& i \frac{d}{dt} X_{f\eta}^{0 \sigma}  = \left( -\mu + \eta\Delta \varepsilon \right) X_{f \eta}^{0 \sigma}
+ \sum_{\delta = \pm x, \pm y} t_{\delta} \eta H_{f \eta \sigma} X_{f+\delta \eta}^{0 \sigma} \nonumber \\
& + & \sum_{\delta} \alpha_{\sigma \delta} H_{f \eta \sigma} X_{f+\delta, \bar{\eta}}^{0 \bar{\sigma}}
+ \sum_{\delta} \Delta_1 \sigma H_{f \eta \sigma} X_{f+\delta, \eta}^{\bar{\sigma} 0},
\end{eqnarray}
where the Hubbard renormalization parameter is $H_{f \eta \sigma} = 1 - \left\langle X_{f\eta}^{\bar{\sigma} \bar{\sigma}} \right\rangle$. As in the case $U=0$ spin-flip correlators $\left\langle X_{f\eta}^{\bar{\sigma} 0} X_{f\eta}^{0 \sigma} \right\rangle$ are neglected.

We use the Zubarev's Green functions, such as
\begin{equation}
\left\langle \left\langle X_{f\eta}^{0 \sigma}(t) | B(t^{\prime}) \right\rangle \right\rangle = -i \Theta(t-t^{\prime}) \left\langle \left\{ X_{f\eta}^{0 \sigma}(t), B(t^{\prime}) \right\} \right\rangle,
\end{equation}
to determine the excitation energy spectrum of the Hubbard fermions and correlators. Here $\Theta(t-t^{\prime})$ is the Heavyside function, $B(t^{\prime})$ is a Hubbard operator of Fermi-type describing creation or annihilation of Hubbard fermion with quantum numbers $\sigma^{\prime}$ and $\eta^{\prime}$ on a site $f^{\prime}$, the braces in the right side denote the anticommutator.  The closed set of equations is obtained for the Fourier transforms of the Green functions $\left\langle \left\langle X_{f\eta}^{0 \sigma} | B \right\rangle \right\rangle_{\omega}$, $\left\langle \left\langle X_{f \bar{\eta}}^{0 \bar{\sigma}} | B \right\rangle \right\rangle_{\omega}$, $\left\langle \left\langle X_{f\eta}^{\bar{\sigma} 0} | B \right\rangle \right\rangle_{\omega}$, $\left\langle \left\langle X_{f\bar{\eta}}^{\sigma 0} | B \right\rangle \right\rangle_{\omega}$. From these equations the spectra both for periodic boundary conditions and for open boundary conditions on a 2D lattice, and edge spectra, when periodic boundary conditions are applied only in one direction of the lattice, are calculated. In the uniform case described in the main text $H_{f \eta \sigma} \equiv H_{\eta} = 1 - n_{\eta}/2$ and $n_{\eta} = \sum_{\sigma} \left\langle X_{f\eta}^{\sigma \sigma} \right\rangle$.

For periodic boundary conditions the self-consistent equation for the electron concentration at the orbital $\eta$ is
\begin{eqnarray}
&& n_{\eta} = \left( 1 - \frac{n_{\eta}}{2} \right)\left( 1 - \sum_k \sum_{j = 1, 2} \frac{(-1)^j}{2E_{jk}\nu_k^2} \times  \right.
\nonumber \\
&\times & \left\{ \xi_{\eta k} \left[ \left( \xi_{\eta k}^2 - \xi_{\bar{\eta}k}^2 \right)/2 + \left( H_{\eta}^2 - H_{\bar{\eta}}^2 \right)/2 \left|\Delta_k \right|^2 + (-1)^j \nu_k^2  \right] \right.
\nonumber \\
&+& \left. \left. \left( \xi_{\eta k} + \xi_{\bar{\eta} k} \right)H_{\eta}H_{\bar{\eta}}\alpha_k^2  \right\}\tanh \left( \frac{E_{j k}}{2T} \right) \right),
\end{eqnarray}
where $\xi_{\eta k} = -\mu + \eta \Delta \varepsilon + \eta H_{\eta} t_k$, $\alpha_{k}^2 = 4\alpha^2 \left( \sin k_x^2 + \sin k_y^2 \right)$, and the HOTSC bulk energy spectrum can be written as
\begin{eqnarray}
E_{jk} & = & \left\{\frac{\xi_{Ak}^2 + \xi_{Bk}^2}{2} + H_{A}H_{B}\alpha_k^2 + \frac{H_A^2 + H_{B}^2}{2}\left|\Delta_k \right|^2 \right.
\nonumber \\
& + & \left. (-1)^j \nu_k^2  \right\}^{1/2},
\end{eqnarray}
and
\begin{eqnarray}
\nu_k^2 & = & \left\{ \left( \xi_{Ak} + \xi_{Bk} \right)^2 \lambda_k^2 + \left|\Delta_k \right|^2 \left[ \frac{H_A^2 - H_{B}^2}{2} \left( \xi_{Ak}^2 - \xi_{Bk}^2 \right) \right. \right.
\nonumber \\
& + & \left. \left.  \left( H_A - H_B \right)^2 H_{A}H_{B}\alpha_k^2 + \left( H_A^2 - H_B^2 \right)^2 \left|\Delta_k \right|^2 \right] \right\}^{1/2}.
\nonumber \\
\end{eqnarray}
Excluding superconducting pairings the bulk energy spectrum of TI is obtained
\begin{eqnarray}
\label{spectr_TI_HI}
&& \varepsilon_{1, 2 k} = \frac{\xi_{A k}+\xi_{B k}}{2} \mp \lambda_{k},
\\
&& \lambda_{k} = \sqrt{\frac{\left(\xi_{A k}-\xi_{B k} \right)^2}{4} + H_A H_B \alpha_{k}^2}.
\nonumber
\end{eqnarray}

\bibliography{hot}

\providecommand{\noopsort}[1]{}\providecommand{\singleletter}[1]{#1}%
\begin{thebibliography}{58}%
\makeatletter
\providecommand \@ifxundefined [1]{%
 \@ifx{#1\undefined}
}%
\providecommand \@ifnum [1]{%
 \ifnum #1\expandafter \@firstoftwo
 \else \expandafter \@secondoftwo
 \fi
}%
\providecommand \@ifx [1]{%
 \ifx #1\expandafter \@firstoftwo
 \else \expandafter \@secondoftwo
 \fi
}%
\providecommand \natexlab [1]{#1}%
\providecommand \enquote  [1]{``#1''}%
\providecommand \bibnamefont  [1]{#1}%
\providecommand \bibfnamefont [1]{#1}%
\providecommand \citenamefont [1]{#1}%
\providecommand \href@noop [0]{\@secondoftwo}%
\providecommand \href [0]{\begingroup \@sanitize@url \@href}%
\providecommand \@href[1]{\@@startlink{#1}\@@href}%
\providecommand \@@href[1]{\endgroup#1\@@endlink}%
\providecommand \@sanitize@url [0]{\catcode `\\12\catcode `\$12\catcode
  `\&12\catcode `\#12\catcode `\^12\catcode `\_12\catcode `\%12\relax}%
\providecommand \@@startlink[1]{}%
\providecommand \@@endlink[0]{}%
\providecommand \url  [0]{\begingroup\@sanitize@url \@url }%
\providecommand \@url [1]{\endgroup\@href {#1}{\urlprefix }}%
\providecommand \urlprefix  [0]{URL }%
\providecommand \Eprint [0]{\href }%
\providecommand \doibase [0]{http://dx.doi.org/}%
\providecommand \selectlanguage [0]{\@gobble}%
\providecommand \bibinfo  [0]{\@secondoftwo}%
\providecommand \bibfield  [0]{\@secondoftwo}%
\providecommand \translation [1]{[#1]}%
\providecommand \BibitemOpen [0]{}%
\providecommand \bibitemStop [0]{}%
\providecommand \bibitemNoStop [0]{.\EOS\space}%
\providecommand \EOS [0]{\spacefactor3000\relax}%
\providecommand \BibitemShut  [1]{\csname bibitem#1\endcsname}%
\let\auto@bib@innerbib\@empty
\bibitem [{\citenamefont {Benalcazar}\ \emph {et~al.}(2017)\citenamefont
  {Benalcazar}, \citenamefont {Bernevig},\ and\ \citenamefont
  {Hughes}}]{benalcazar-17}%
  \BibitemOpen
  \bibfield  {author} {\bibinfo {author} {\bibfnamefont {W.~A.}\ \bibnamefont
  {Benalcazar}}, \bibinfo {author} {\bibfnamefont {B.~A.}\ \bibnamefont
  {Bernevig}}, \ and\ \bibinfo {author} {\bibfnamefont {T.~L.}\ \bibnamefont
  {Hughes}},\ }\href {\doibase 10.1126/science.aah6442} {\bibfield  {journal}
  {\bibinfo  {journal} {Science}\ }\textbf {\bibinfo {volume} {357}},\ \bibinfo
  {pages} {61} (\bibinfo {year} {2017})}\BibitemShut {NoStop}%
\bibitem [{\citenamefont {Langbehn}\ \emph {et~al.}(2017)\citenamefont
  {Langbehn}, \citenamefont {Peng}, \citenamefont {Trifunovic}, \citenamefont
  {von Oppen},\ and\ \citenamefont {Brouwer}}]{Langbehn-17}%
  \BibitemOpen
  \bibfield  {author} {\bibinfo {author} {\bibfnamefont {J.}~\bibnamefont
  {Langbehn}}, \bibinfo {author} {\bibfnamefont {Y.}~\bibnamefont {Peng}},
  \bibinfo {author} {\bibfnamefont {L.}~\bibnamefont {Trifunovic}}, \bibinfo
  {author} {\bibfnamefont {F.}~\bibnamefont {von Oppen}}, \ and\ \bibinfo
  {author} {\bibfnamefont {P.~W.}\ \bibnamefont {Brouwer}},\ }\href {\doibase
  10.1103/PhysRevLett.119.246401} {\bibfield  {journal} {\bibinfo  {journal}
  {Phys.\ Rev.\ Lett.}\ }\textbf {\bibinfo {volume} {119}},\ \bibinfo {pages}
  {246401} (\bibinfo {year} {2017})}\BibitemShut {NoStop}%
\bibitem [{\citenamefont {Zlotnikov}\ \emph {et~al.}(2021)\citenamefont
  {Zlotnikov}, \citenamefont {Shustin},\ and\ \citenamefont
  {Fedoseev}}]{zlotnikov-21}%
  \BibitemOpen
  \bibfield  {author} {\bibinfo {author} {\bibfnamefont {A.~O.}\ \bibnamefont
  {Zlotnikov}}, \bibinfo {author} {\bibfnamefont {M.~S.}\ \bibnamefont
  {Shustin}}, \ and\ \bibinfo {author} {\bibfnamefont {A.~D.}\ \bibnamefont
  {Fedoseev}},\ }\href {\doibase 10.1007/s10948-021-06029-z} {\bibfield
  {journal} {\bibinfo  {journal} {J Supercond Nov Magn}\ }\textbf {\bibinfo
  {volume} {34}},\ \bibinfo {pages} {3053} (\bibinfo {year}
  {2021})}\BibitemShut {NoStop}%
\bibitem [{\citenamefont {Volovik}(2010)}]{volovik-10}%
  \BibitemOpen
  \bibfield  {author} {\bibinfo {author} {\bibfnamefont {G.~E.}\ \bibnamefont
  {Volovik}},\ }\href {\doibase 10.1134/S0021364010040090} {\bibfield
  {journal} {\bibinfo  {journal} {Jetp Lett.}\ }\textbf {\bibinfo {volume}
  {91}},\ \bibinfo {pages} {201} (\bibinfo {year} {2010})}\BibitemShut
  {NoStop}%
\bibitem [{\citenamefont {Ivanov}(2001)}]{ivanov-01}%
  \BibitemOpen
  \bibfield  {author} {\bibinfo {author} {\bibfnamefont {D.~A.}\ \bibnamefont
  {Ivanov}},\ }\href {\doibase 10.1103/PhysRevLett.86.268} {\bibfield
  {journal} {\bibinfo  {journal} {Phys.\ Rev.\ Lett.}\ }\textbf {\bibinfo
  {volume} {86}},\ \bibinfo {pages} {268} (\bibinfo {year} {2001})}\BibitemShut
  {NoStop}%
\bibitem [{\citenamefont {Alicea}\ \emph {et~al.}(2011)\citenamefont {Alicea},
  \citenamefont {Oreg}, \citenamefont {Refael}, \citenamefont {von Oppen},\
  and\ \citenamefont {Fisher}}]{alicea-11}%
  \BibitemOpen
  \bibfield  {author} {\bibinfo {author} {\bibfnamefont {J.}~\bibnamefont
  {Alicea}}, \bibinfo {author} {\bibfnamefont {Y.}~\bibnamefont {Oreg}},
  \bibinfo {author} {\bibfnamefont {G.}~\bibnamefont {Refael}}, \bibinfo
  {author} {\bibfnamefont {F.}~\bibnamefont {von Oppen}}, \ and\ \bibinfo
  {author} {\bibfnamefont {M.~P.~A.}\ \bibnamefont {Fisher}},\ }\href {\doibase
  10.1038/NPHYS1915} {\bibfield  {journal} {\bibinfo  {journal} {Nat.\ Phys.}\
  }\textbf {\bibinfo {volume} {7}},\ \bibinfo {pages} {412} (\bibinfo {year}
  {2011})}\BibitemShut {NoStop}%
\bibitem [{\citenamefont {Kitaev}(2001)}]{kitaev-01}%
  \BibitemOpen
  \bibfield  {author} {\bibinfo {author} {\bibfnamefont {A.~Y.}\ \bibnamefont
  {Kitaev}},\ }\href {\doibase http://dx.doi.org/10.1070/1063-7869/44/10S/S29}
  {\bibfield  {journal} {\bibinfo  {journal} {Phys.\ Usp.}\ }\textbf {\bibinfo
  {volume} {44}},\ \bibinfo {pages} {131} (\bibinfo {year} {2001})}\BibitemShut
  {NoStop}%
\bibitem [{\citenamefont {Lutchyn}\ \emph {et~al.}(2010)\citenamefont
  {Lutchyn}, \citenamefont {Sau},\ and\ \citenamefont {Sarma}}]{lutchyn-10}%
  \BibitemOpen
  \bibfield  {author} {\bibinfo {author} {\bibfnamefont {R.~M.}\ \bibnamefont
  {Lutchyn}}, \bibinfo {author} {\bibfnamefont {J.~D.}\ \bibnamefont {Sau}}, \
  and\ \bibinfo {author} {\bibfnamefont {S.~D.}\ \bibnamefont {Sarma}},\ }\href
  {\doibase 10.1103/PhysRevLett.105.077001} {\bibfield  {journal} {\bibinfo
  {journal} {Phys.\ Rev.\ Lett.}\ }\textbf {\bibinfo {volume} {105}},\ \bibinfo
  {pages} {077001} (\bibinfo {year} {2010})}\BibitemShut {NoStop}%
\bibitem [{\citenamefont {Oreg}\ \emph {et~al.}(2010)\citenamefont {Oreg},
  \citenamefont {Refael},\ and\ \citenamefont {von Oppen}}]{oreg-10}%
  \BibitemOpen
  \bibfield  {author} {\bibinfo {author} {\bibfnamefont {Y.}~\bibnamefont
  {Oreg}}, \bibinfo {author} {\bibfnamefont {G.}~\bibnamefont {Refael}}, \ and\
  \bibinfo {author} {\bibfnamefont {F.}~\bibnamefont {von Oppen}},\ }\href
  {\doibase 10.1103/PhysRevLett.105.177002} {\bibfield  {journal} {\bibinfo
  {journal} {Phys.\ Rev.\ Lett.}\ }\textbf {\bibinfo {volume} {105}},\ \bibinfo
  {pages} {177002} (\bibinfo {year} {2010})}\BibitemShut {NoStop}%
\bibitem [{\citenamefont {Potter}\ and\ \citenamefont {Lee}(2010)}]{Potter-10}%
  \BibitemOpen
  \bibfield  {author} {\bibinfo {author} {\bibfnamefont {A.~C.}\ \bibnamefont
  {Potter}}\ and\ \bibinfo {author} {\bibfnamefont {P.~A.}\ \bibnamefont
  {Lee}},\ }\href {\doibase 10.1103/PhysRevLett.105.227003} {\bibfield
  {journal} {\bibinfo  {journal} {Phys.\ Rev.\ Lett.}\ }\textbf {\bibinfo
  {volume} {105}},\ \bibinfo {pages} {227003} (\bibinfo {year}
  {2010})}\BibitemShut {NoStop}%
\bibitem [{\citenamefont {Sedlmayr}\ \emph {et~al.}(2016)\citenamefont
  {Sedlmayr}, \citenamefont {Aguiar-Hualde},\ and\ \citenamefont
  {Bena}}]{Sedlmayr-16}%
  \BibitemOpen
  \bibfield  {author} {\bibinfo {author} {\bibfnamefont {N.}~\bibnamefont
  {Sedlmayr}}, \bibinfo {author} {\bibfnamefont {J.~M.}\ \bibnamefont
  {Aguiar-Hualde}}, \ and\ \bibinfo {author} {\bibfnamefont {C.}~\bibnamefont
  {Bena}},\ }\href {\doibase 10.1103/PhysRevB.93.155425} {\bibfield  {journal}
  {\bibinfo  {journal} {Phys.\ Rev.\ B}\ }\textbf {\bibinfo {volume} {93}},\
  \bibinfo {pages} {155425} (\bibinfo {year} {2016})}\BibitemShut {NoStop}%
\bibitem [{\citenamefont {Nayak}\ \emph {et~al.}(2008)\citenamefont {Nayak},
  \citenamefont {Simon}, \citenamefont {Stern}, \citenamefont {Freedman},\ and\
  \citenamefont {DasSarma}}]{Nayak-08}%
  \BibitemOpen
  \bibfield  {author} {\bibinfo {author} {\bibfnamefont {C.}~\bibnamefont
  {Nayak}}, \bibinfo {author} {\bibfnamefont {S.~H.}\ \bibnamefont {Simon}},
  \bibinfo {author} {\bibfnamefont {A.}~\bibnamefont {Stern}}, \bibinfo
  {author} {\bibfnamefont {M.}~\bibnamefont {Freedman}}, \ and\ \bibinfo
  {author} {\bibfnamefont {S.}~\bibnamefont {DasSarma}},\ }\href {\doibase
  10.1103/RevModPhys.80.1083} {\bibfield  {journal} {\bibinfo  {journal} {Rev.\
  Mod.\ Phys.}\ }\textbf {\bibinfo {volume} {80}},\ \bibinfo {pages} {1083}
  (\bibinfo {year} {2008})}\BibitemShut {NoStop}%
\bibitem [{\citenamefont {Cheng}\ \emph {et~al.}(2016)\citenamefont {Cheng},
  \citenamefont {He},\ and\ \citenamefont {Kou}}]{Cheng-16}%
  \BibitemOpen
  \bibfield  {author} {\bibinfo {author} {\bibfnamefont {Q.}~\bibnamefont
  {Cheng}}, \bibinfo {author} {\bibfnamefont {J.}~\bibnamefont {He}}, \ and\
  \bibinfo {author} {\bibfnamefont {S.~P.}\ \bibnamefont {Kou}},\ }\href
  {\doibase 10.1016/j.physleta.2015.11.030} {\bibfield  {journal} {\bibinfo
  {journal} {Phys.\ Lett.\ A}\ }\textbf {\bibinfo {volume} {380}},\ \bibinfo
  {pages} {779} (\bibinfo {year} {2016})}\BibitemShut {NoStop}%
\bibitem [{\citenamefont {Harper}\ \emph {et~al.}(2019)\citenamefont {Harper},
  \citenamefont {Pushp},\ and\ \citenamefont {Roy}}]{Harper-19}%
  \BibitemOpen
  \bibfield  {author} {\bibinfo {author} {\bibfnamefont {F.}~\bibnamefont
  {Harper}}, \bibinfo {author} {\bibfnamefont {A.}~\bibnamefont {Pushp}}, \
  and\ \bibinfo {author} {\bibfnamefont {R.}~\bibnamefont {Roy}},\ }\href
  {\doibase 10.1103/PhysRevResearch.1.033207} {\bibfield  {journal} {\bibinfo
  {journal} {Phys.\ Rev.\ Research}\ }\textbf {\bibinfo {volume} {1}},\
  \bibinfo {pages} {033207} (\bibinfo {year} {2019})}\BibitemShut {NoStop}%
\bibitem [{\citenamefont {Zhou}\ \emph {et~al.}(2020)\citenamefont {Zhou},
  \citenamefont {Dartiailh}, \citenamefont {Mayer}, \citenamefont {Han},
  \citenamefont {Matos-Abiague}, \citenamefont {Shabani},\ and\ \citenamefont
  {Zutic}}]{Zhou-20}%
  \BibitemOpen
  \bibfield  {author} {\bibinfo {author} {\bibfnamefont {T.}~\bibnamefont
  {Zhou}}, \bibinfo {author} {\bibfnamefont {M.~C.}\ \bibnamefont {Dartiailh}},
  \bibinfo {author} {\bibfnamefont {W.}~\bibnamefont {Mayer}}, \bibinfo
  {author} {\bibfnamefont {J.~E.}\ \bibnamefont {Han}}, \bibinfo {author}
  {\bibfnamefont {A.}~\bibnamefont {Matos-Abiague}}, \bibinfo {author}
  {\bibfnamefont {J.}~\bibnamefont {Shabani}}, \ and\ \bibinfo {author}
  {\bibfnamefont {I.}~\bibnamefont {Zutic}},\ }\href {\doibase
  10.1103/PhysRevLett.124.137001} {\bibfield  {journal} {\bibinfo  {journal}
  {Phys.\ Rev.\ Lett.}\ }\textbf {\bibinfo {volume} {124}},\ \bibinfo {pages}
  {137001} (\bibinfo {year} {2020})}\BibitemShut {NoStop}%
\bibitem [{\citenamefont {Zhu}(2018)}]{Zhu-18}%
  \BibitemOpen
  \bibfield  {author} {\bibinfo {author} {\bibfnamefont {X.}~\bibnamefont
  {Zhu}},\ }\href {\doibase 10.1103/PhysRevB.97.205134} {\bibfield  {journal}
  {\bibinfo  {journal} {Phys.\ Rev.\ B}\ }\textbf {\bibinfo {volume} {97}},\
  \bibinfo {pages} {205134} (\bibinfo {year} {2018})}\BibitemShut {NoStop}%
\bibitem [{\citenamefont {Franca}\ \emph {et~al.}(2019)\citenamefont {Franca},
  \citenamefont {Efremov},\ and\ \citenamefont {Fulga}}]{Franca-19}%
  \BibitemOpen
  \bibfield  {author} {\bibinfo {author} {\bibfnamefont {S.}~\bibnamefont
  {Franca}}, \bibinfo {author} {\bibfnamefont {D.~V.}\ \bibnamefont {Efremov}},
  \ and\ \bibinfo {author} {\bibfnamefont {I.~C.}\ \bibnamefont {Fulga}},\
  }\href {\doibase 10.1103/PhysRevB.100.075415} {\bibfield  {journal} {\bibinfo
   {journal} {Phys. Rev. B}\ }\textbf {\bibinfo {volume} {100}},\ \bibinfo
  {pages} {075415} (\bibinfo {year} {2019})}\BibitemShut {NoStop}%
\bibitem [{\citenamefont {Wu}\ \emph {et~al.}(2020)\citenamefont {Wu},
  \citenamefont {Hou}, \citenamefont {Li}, \citenamefont {Luo}, \citenamefont
  {Shi},\ and\ \citenamefont {Zhang}}]{Wu-20}%
  \BibitemOpen
  \bibfield  {author} {\bibinfo {author} {\bibfnamefont {Y.-J.}\ \bibnamefont
  {Wu}}, \bibinfo {author} {\bibfnamefont {J.}~\bibnamefont {Hou}}, \bibinfo
  {author} {\bibfnamefont {Y.-M.}\ \bibnamefont {Li}}, \bibinfo {author}
  {\bibfnamefont {X.-W.}\ \bibnamefont {Luo}}, \bibinfo {author} {\bibfnamefont
  {X.}~\bibnamefont {Shi}}, \ and\ \bibinfo {author} {\bibfnamefont
  {C.}~\bibnamefont {Zhang}},\ }\href {\doibase 10.1103/PhysRevLett.124.227001}
  {\bibfield  {journal} {\bibinfo  {journal} {Phys. Rev. Lett.}\ }\textbf
  {\bibinfo {volume} {124}},\ \bibinfo {pages} {227001} (\bibinfo {year}
  {2020})}\BibitemShut {NoStop}%
\bibitem [{\citenamefont {Plekhanov}\ \emph {et~al.}(2021)\citenamefont
  {Plekhanov}, \citenamefont {M\"uller}, \citenamefont {Volpez}, \citenamefont
  {Kennes}, \citenamefont {Schoeller}, \citenamefont {Loss},\ and\
  \citenamefont {Klinovaja}}]{Plekhanov-21}%
  \BibitemOpen
  \bibfield  {author} {\bibinfo {author} {\bibfnamefont {K.}~\bibnamefont
  {Plekhanov}}, \bibinfo {author} {\bibfnamefont {N.}~\bibnamefont {M\"uller}},
  \bibinfo {author} {\bibfnamefont {Y.}~\bibnamefont {Volpez}}, \bibinfo
  {author} {\bibfnamefont {D.~M.}\ \bibnamefont {Kennes}}, \bibinfo {author}
  {\bibfnamefont {H.}~\bibnamefont {Schoeller}}, \bibinfo {author}
  {\bibfnamefont {D.}~\bibnamefont {Loss}}, \ and\ \bibinfo {author}
  {\bibfnamefont {J.}~\bibnamefont {Klinovaja}},\ }\href {\doibase
  10.1103/PhysRevB.103.L041401} {\bibfield  {journal} {\bibinfo  {journal}
  {Phys. Rev. B}\ }\textbf {\bibinfo {volume} {103}},\ \bibinfo {pages}
  {L041401} (\bibinfo {year} {2021})}\BibitemShut {NoStop}%
\bibitem [{\citenamefont {Pahomi}\ \emph {et~al.}(2020)\citenamefont {Pahomi},
  \citenamefont {Sigrist},\ and\ \citenamefont {Soluyanov}}]{pahomi-20}%
  \BibitemOpen
  \bibfield  {author} {\bibinfo {author} {\bibfnamefont {T.~E.}\ \bibnamefont
  {Pahomi}}, \bibinfo {author} {\bibfnamefont {M.}~\bibnamefont {Sigrist}}, \
  and\ \bibinfo {author} {\bibfnamefont {A.~A.}\ \bibnamefont {Soluyanov}},\
  }\href {\doibase 10.1103/PhysRevResearch.2.032068} {\bibfield  {journal}
  {\bibinfo  {journal} {Phys.\ Rev.\ Research}\ }\textbf {\bibinfo {volume}
  {2}},\ \bibinfo {pages} {032068(R)} (\bibinfo {year} {2020})}\BibitemShut
  {NoStop}%
\bibitem [{\citenamefont {Zhang}\ \emph
  {et~al.}(2020{\natexlab{a}})\citenamefont {Zhang}, \citenamefont {Rui},
  \citenamefont {Calzona}, \citenamefont {Choi}, \citenamefont {Schnyder},\
  and\ \citenamefont {Trauzettel}}]{zhang-20PRR}%
  \BibitemOpen
  \bibfield  {author} {\bibinfo {author} {\bibfnamefont {S.~B.}\ \bibnamefont
  {Zhang}}, \bibinfo {author} {\bibfnamefont {W.~B.}\ \bibnamefont {Rui}},
  \bibinfo {author} {\bibfnamefont {A.}~\bibnamefont {Calzona}}, \bibinfo
  {author} {\bibfnamefont {S.~J.}\ \bibnamefont {Choi}}, \bibinfo {author}
  {\bibfnamefont {A.~P.}\ \bibnamefont {Schnyder}}, \ and\ \bibinfo {author}
  {\bibfnamefont {B.}~\bibnamefont {Trauzettel}},\ }\href {\doibase
  10.1103/PhysRevResearch.2.043025} {\bibfield  {journal} {\bibinfo  {journal}
  {Phys.\ Rev.\ Research}\ }\textbf {\bibinfo {volume} {2}},\ \bibinfo {pages}
  {043025} (\bibinfo {year} {2020}{\natexlab{a}})}\BibitemShut {NoStop}%
\bibitem [{\citenamefont {Zhang}\ \emph
  {et~al.}(2020{\natexlab{b}})\citenamefont {Zhang}, \citenamefont {Calzona},\
  and\ \citenamefont {Trauzettel}}]{zhang-20PRB}%
  \BibitemOpen
  \bibfield  {author} {\bibinfo {author} {\bibfnamefont {S.~B.}\ \bibnamefont
  {Zhang}}, \bibinfo {author} {\bibfnamefont {A.}~\bibnamefont {Calzona}}, \
  and\ \bibinfo {author} {\bibfnamefont {B.}~\bibnamefont {Trauzettel}},\
  }\href {\doibase 10.1103/PhysRevB.102.100503} {\bibfield  {journal} {\bibinfo
   {journal} {Phys.\ Rev.\ B}\ }\textbf {\bibinfo {volume} {102}},\ \bibinfo
  {pages} {100503(R)} (\bibinfo {year} {2020}{\natexlab{b}})}\BibitemShut
  {NoStop}%
\bibitem [{\citenamefont {Hsu}\ \emph {et~al.}(2020)\citenamefont {Hsu},
  \citenamefont {Cole}, \citenamefont {Zhang},\ and\ \citenamefont
  {Sau}}]{hsu-20}%
  \BibitemOpen
  \bibfield  {author} {\bibinfo {author} {\bibfnamefont {Y.~T.}\ \bibnamefont
  {Hsu}}, \bibinfo {author} {\bibfnamefont {W.~S.}\ \bibnamefont {Cole}},
  \bibinfo {author} {\bibfnamefont {R.~X.}\ \bibnamefont {Zhang}}, \ and\
  \bibinfo {author} {\bibfnamefont {J.~D.}\ \bibnamefont {Sau}},\ }\href
  {\doibase 10.1103/PhysRevLett.125.097001} {\bibfield  {journal} {\bibinfo
  {journal} {Phys.\ Rev.\ Lett.}\ }\textbf {\bibinfo {volume} {125}},\ \bibinfo
  {pages} {097001} (\bibinfo {year} {2020})}\BibitemShut {NoStop}%
\bibitem [{\citenamefont {Kheirkhah}\ \emph {et~al.}(2020)\citenamefont
  {Kheirkhah}, \citenamefont {Yan}, \citenamefont {Nagai},\ and\ \citenamefont
  {Marsiglio}}]{kheirkhah-20}%
  \BibitemOpen
  \bibfield  {author} {\bibinfo {author} {\bibfnamefont {M.}~\bibnamefont
  {Kheirkhah}}, \bibinfo {author} {\bibfnamefont {Z.}~\bibnamefont {Yan}},
  \bibinfo {author} {\bibfnamefont {Y.}~\bibnamefont {Nagai}}, \ and\ \bibinfo
  {author} {\bibfnamefont {F.}~\bibnamefont {Marsiglio}},\ }\href {\doibase
  10.1103/PhysRevLett.125.017001} {\bibfield  {journal} {\bibinfo  {journal}
  {Phys.\ Rev.\ Lett.}\ }\textbf {\bibinfo {volume} {125}},\ \bibinfo {pages}
  {017001} (\bibinfo {year} {2020})}\BibitemShut {NoStop}%
\bibitem [{\citenamefont {Li}\ \emph {et~al.}(2022)\citenamefont {Li},
  \citenamefont {Geier}, \citenamefont {Ingham},\ and\ \citenamefont
  {Scammell}}]{li-22}%
  \BibitemOpen
  \bibfield  {author} {\bibinfo {author} {\bibfnamefont {T.}~\bibnamefont
  {Li}}, \bibinfo {author} {\bibfnamefont {M.}~\bibnamefont {Geier}}, \bibinfo
  {author} {\bibfnamefont {J.}~\bibnamefont {Ingham}}, \ and\ \bibinfo {author}
  {\bibfnamefont {H.~D.}\ \bibnamefont {Scammell}},\ }\href@noop {} {\bibfield
  {journal} {\bibinfo  {journal} {2D Mater.}\ }\textbf {\bibinfo {volume}
  {9}},\ \bibinfo {pages} {015031} (\bibinfo {year} {2022})}\BibitemShut
  {NoStop}%
\bibitem [{\citenamefont {Stoudenmire}\ \emph {et~al.}(2011)\citenamefont
  {Stoudenmire}, \citenamefont {Alicea}, \citenamefont {Starykh},\ and\
  \citenamefont {Fisher}}]{stoudenmire-11}%
  \BibitemOpen
  \bibfield  {author} {\bibinfo {author} {\bibfnamefont {E.}~\bibnamefont
  {Stoudenmire}}, \bibinfo {author} {\bibfnamefont {J.}~\bibnamefont {Alicea}},
  \bibinfo {author} {\bibfnamefont {O.}~\bibnamefont {Starykh}}, \ and\
  \bibinfo {author} {\bibfnamefont {M.}~\bibnamefont {Fisher}},\ }\href@noop {}
  {\bibfield  {journal} {\bibinfo  {journal} {Phys.\ Rev.\ B}\ }\textbf
  {\bibinfo {volume} {84}},\ \bibinfo {pages} {014503} (\bibinfo {year}
  {2011})}\BibitemShut {NoStop}%
\bibitem [{\citenamefont {Thomale}\ \emph {et~al.}(2013)\citenamefont
  {Thomale}, \citenamefont {Rachel},\ and\ \citenamefont
  {Schmitteckert}}]{thomale-13}%
  \BibitemOpen
  \bibfield  {author} {\bibinfo {author} {\bibfnamefont {R.}~\bibnamefont
  {Thomale}}, \bibinfo {author} {\bibfnamefont {S.}~\bibnamefont {Rachel}}, \
  and\ \bibinfo {author} {\bibfnamefont {P.}~\bibnamefont {Schmitteckert}},\
  }\href@noop {} {\bibfield  {journal} {\bibinfo  {journal} {Phys.\ Rev.\ B}\
  }\textbf {\bibinfo {volume} {88}},\ \bibinfo {pages} {161103(R)} (\bibinfo
  {year} {2013})}\BibitemShut {NoStop}%
\bibitem [{\citenamefont {Katsura}\ \emph {et~al.}(2015)\citenamefont
  {Katsura}, \citenamefont {Schuricht},\ and\ \citenamefont
  {Takahashi}}]{katsura-15}%
  \BibitemOpen
  \bibfield  {author} {\bibinfo {author} {\bibfnamefont {H.}~\bibnamefont
  {Katsura}}, \bibinfo {author} {\bibfnamefont {D.}~\bibnamefont {Schuricht}},
  \ and\ \bibinfo {author} {\bibfnamefont {M.}~\bibnamefont {Takahashi}},\
  }\href@noop {} {\bibfield  {journal} {\bibinfo  {journal} {Phys.\ Rev.\ B}\
  }\textbf {\bibinfo {volume} {92}},\ \bibinfo {pages} {115137} (\bibinfo
  {year} {2015})}\BibitemShut {NoStop}%
\bibitem [{\citenamefont {Aksenov}\ \emph {et~al.}(2020)\citenamefont
  {Aksenov}, \citenamefont {Zlotnikov},\ and\ \citenamefont
  {Shustin}}]{aksenov-20}%
  \BibitemOpen
  \bibfield  {author} {\bibinfo {author} {\bibfnamefont {S.~V.}\ \bibnamefont
  {Aksenov}}, \bibinfo {author} {\bibfnamefont {A.~O.}\ \bibnamefont
  {Zlotnikov}}, \ and\ \bibinfo {author} {\bibfnamefont {M.~S.}\ \bibnamefont
  {Shustin}},\ }\href {\doibase 10.1103/PhysRevB.101.125431} {\bibfield
  {journal} {\bibinfo  {journal} {Phys. Rev. B}\ }\textbf {\bibinfo {volume}
  {101}},\ \bibinfo {pages} {125431} (\bibinfo {year} {2020})}\BibitemShut
  {NoStop}%
\bibitem [{\citenamefont {Kudo}\ \emph {et~al.}(2019)\citenamefont {Kudo},
  \citenamefont {Yoshida},\ and\ \citenamefont {Hatsugai}}]{kudo-19}%
  \BibitemOpen
  \bibfield  {author} {\bibinfo {author} {\bibfnamefont {K.}~\bibnamefont
  {Kudo}}, \bibinfo {author} {\bibfnamefont {T.}~\bibnamefont {Yoshida}}, \
  and\ \bibinfo {author} {\bibfnamefont {Y.}~\bibnamefont {Hatsugai}},\ }\href
  {\doibase 10.1103/PhysRevLett.123.196402} {\bibfield  {journal} {\bibinfo
  {journal} {Phys.\ Rev.\ Lett.}\ }\textbf {\bibinfo {volume} {123}},\ \bibinfo
  {pages} {196402} (\bibinfo {year} {2019})}\BibitemShut {NoStop}%
\bibitem [{\citenamefont {Otsuka}\ \emph {et~al.}(2021)\citenamefont {Otsuka},
  \citenamefont {Yoshida}, \citenamefont {Kudo}, \citenamefont {Yunoki},\ and\
  \citenamefont {Hatsugai}}]{otsuka-21}%
  \BibitemOpen
  \bibfield  {author} {\bibinfo {author} {\bibfnamefont {Y.}~\bibnamefont
  {Otsuka}}, \bibinfo {author} {\bibfnamefont {T.}~\bibnamefont {Yoshida}},
  \bibinfo {author} {\bibfnamefont {K.}~\bibnamefont {Kudo}}, \bibinfo {author}
  {\bibfnamefont {S.}~\bibnamefont {Yunoki}}, \ and\ \bibinfo {author}
  {\bibfnamefont {Y.}~\bibnamefont {Hatsugai}},\ }\href@noop {} {\bibfield
  {journal} {\bibinfo  {journal} {Sci.\ Rep.}\ }\textbf {\bibinfo {volume}
  {11}},\ \bibinfo {pages} {20270} (\bibinfo {year} {2021})}\BibitemShut
  {NoStop}%
\bibitem [{\citenamefont {Zhao}\ \emph {et~al.}(2021)\citenamefont {Zhao},
  \citenamefont {Qiang}, \citenamefont {Lu},\ and\ \citenamefont
  {Xie}}]{zhao-21}%
  \BibitemOpen
  \bibfield  {author} {\bibinfo {author} {\bibfnamefont {P.-L.}\ \bibnamefont
  {Zhao}}, \bibinfo {author} {\bibfnamefont {X.-B.}\ \bibnamefont {Qiang}},
  \bibinfo {author} {\bibfnamefont {H.-Z.}\ \bibnamefont {Lu}}, \ and\ \bibinfo
  {author} {\bibfnamefont {X.~C.}\ \bibnamefont {Xie}},\ }\href {\doibase
  10.1103/PhysRevLett.127.176601} {\bibfield  {journal} {\bibinfo  {journal}
  {Phys.\ Rev.\ Lett.}\ }\textbf {\bibinfo {volume} {127}},\ \bibinfo {pages}
  {176601} (\bibinfo {year} {2021})}\BibitemShut {NoStop}%
\bibitem [{\citenamefont {Hassan}\ \emph {et~al.}(2019)\citenamefont {Hassan},
  \citenamefont {Kunst}, \citenamefont {Moritz}, \citenamefont {Andler},
  \citenamefont {Bergholtz},\ and\ \citenamefont {Bourennane}}]{el_hassan-19}%
  \BibitemOpen
  \bibfield  {author} {\bibinfo {author} {\bibfnamefont {A.~E.}\ \bibnamefont
  {Hassan}}, \bibinfo {author} {\bibfnamefont {F.~K.}\ \bibnamefont {Kunst}},
  \bibinfo {author} {\bibfnamefont {A.}~\bibnamefont {Moritz}}, \bibinfo
  {author} {\bibfnamefont {G.}~\bibnamefont {Andler}}, \bibinfo {author}
  {\bibfnamefont {E.~J.}\ \bibnamefont {Bergholtz}}, \ and\ \bibinfo {author}
  {\bibfnamefont {M.}~\bibnamefont {Bourennane}},\ }\href@noop {} {\bibfield
  {journal} {\bibinfo  {journal} {Nat.\ Photonics}\ }\textbf {\bibinfo {volume}
  {13}},\ \bibinfo {pages} {697} (\bibinfo {year} {2019})}\BibitemShut
  {NoStop}%
\bibitem [{\citenamefont {Ni}\ \emph {et~al.}(2019)\citenamefont {Ni},
  \citenamefont {Weiner}, \citenamefont {Alu},\ and\ \citenamefont
  {Khanikaev}}]{ni-19}%
  \BibitemOpen
  \bibfield  {author} {\bibinfo {author} {\bibfnamefont {X.}~\bibnamefont
  {Ni}}, \bibinfo {author} {\bibfnamefont {M.}~\bibnamefont {Weiner}}, \bibinfo
  {author} {\bibfnamefont {A.}~\bibnamefont {Alu}}, \ and\ \bibinfo {author}
  {\bibfnamefont {A.~B.}\ \bibnamefont {Khanikaev}},\ }\href@noop {} {\bibfield
   {journal} {\bibinfo  {journal} {Nat.\ Mater.}\ }\textbf {\bibinfo {volume}
  {18}},\ \bibinfo {pages} {113} (\bibinfo {year} {2019})}\BibitemShut
  {NoStop}%
\bibitem [{\citenamefont {Imhof}\ \emph {et~al.}(2018)\citenamefont {Imhof},
  \citenamefont {Berger}, \citenamefont {Bayer}, \citenamefont {Brehm},
  \citenamefont {Molenkamp}, \citenamefont {Kiessling}, \citenamefont
  {Schindler}, \citenamefont {Lee}, \citenamefont {Greiter}, \citenamefont
  {Neupert},\ and\ \citenamefont {Thomalem}}]{imhof-18}%
  \BibitemOpen
  \bibfield  {author} {\bibinfo {author} {\bibfnamefont {S.}~\bibnamefont
  {Imhof}}, \bibinfo {author} {\bibfnamefont {C.}~\bibnamefont {Berger}},
  \bibinfo {author} {\bibfnamefont {F.}~\bibnamefont {Bayer}}, \bibinfo
  {author} {\bibfnamefont {J.}~\bibnamefont {Brehm}}, \bibinfo {author}
  {\bibfnamefont {L.~W.}\ \bibnamefont {Molenkamp}}, \bibinfo {author}
  {\bibfnamefont {T.}~\bibnamefont {Kiessling}}, \bibinfo {author}
  {\bibfnamefont {F.}~\bibnamefont {Schindler}}, \bibinfo {author}
  {\bibfnamefont {C.~H.}\ \bibnamefont {Lee}}, \bibinfo {author} {\bibfnamefont
  {M.}~\bibnamefont {Greiter}}, \bibinfo {author} {\bibfnamefont
  {T.}~\bibnamefont {Neupert}}, \ and\ \bibinfo {author} {\bibfnamefont
  {R.}~\bibnamefont {Thomalem}},\ }\href@noop {} {\bibfield  {journal}
  {\bibinfo  {journal} {Nat.\ Phys.}\ }\textbf {\bibinfo {volume} {14}},\
  \bibinfo {pages} {925} (\bibinfo {year} {2018})}\BibitemShut {NoStop}%
\bibitem [{\citenamefont {Serra-Garcia}\ \emph {et~al.}(2019)\citenamefont
  {Serra-Garcia}, \citenamefont {Susstrunk},\ and\ \citenamefont
  {Huber}}]{serra-garcia-19}%
  \BibitemOpen
  \bibfield  {author} {\bibinfo {author} {\bibfnamefont {M.}~\bibnamefont
  {Serra-Garcia}}, \bibinfo {author} {\bibfnamefont {R.}~\bibnamefont
  {Susstrunk}}, \ and\ \bibinfo {author} {\bibfnamefont {S.~D.}\ \bibnamefont
  {Huber}},\ }\href@noop {} {\bibfield  {journal} {\bibinfo  {journal} {Phys.\
  Rev.\ B}\ }\textbf {\bibinfo {volume} {99}},\ \bibinfo {pages} {020304}
  (\bibinfo {year} {2019})}\BibitemShut {NoStop}%
\bibitem [{\citenamefont {Schindler}\ \emph {et~al.}(2018)\citenamefont
  {Schindler}, \citenamefont {Wang}, \citenamefont {Vergniory}, \citenamefont
  {Cook}, \citenamefont {Murani}, \citenamefont {Sengupta}, \citenamefont
  {Kasumov}, \citenamefont {Deblock}, \citenamefont {Jeon}, \citenamefont
  {Drozdov}, \citenamefont {Bouchiat}, \citenamefont {Gueron}, \citenamefont
  {Yazdani}, \citenamefont {Bernevig},\ and\ \citenamefont
  {Neupert}}]{Schindler-18NP}%
  \BibitemOpen
  \bibfield  {author} {\bibinfo {author} {\bibfnamefont {F.}~\bibnamefont
  {Schindler}}, \bibinfo {author} {\bibfnamefont {Z.}~\bibnamefont {Wang}},
  \bibinfo {author} {\bibfnamefont {M.~G.}\ \bibnamefont {Vergniory}}, \bibinfo
  {author} {\bibfnamefont {A.~M.}\ \bibnamefont {Cook}}, \bibinfo {author}
  {\bibfnamefont {A.}~\bibnamefont {Murani}}, \bibinfo {author} {\bibfnamefont
  {S.}~\bibnamefont {Sengupta}}, \bibinfo {author} {\bibfnamefont {A.~Y.}\
  \bibnamefont {Kasumov}}, \bibinfo {author} {\bibfnamefont {R.}~\bibnamefont
  {Deblock}}, \bibinfo {author} {\bibfnamefont {S.}~\bibnamefont {Jeon}},
  \bibinfo {author} {\bibfnamefont {I.}~\bibnamefont {Drozdov}}, \bibinfo
  {author} {\bibfnamefont {H.}~\bibnamefont {Bouchiat}}, \bibinfo {author}
  {\bibfnamefont {S.}~\bibnamefont {Gueron}}, \bibinfo {author} {\bibfnamefont
  {A.}~\bibnamefont {Yazdani}}, \bibinfo {author} {\bibfnamefont {B.~A.}\
  \bibnamefont {Bernevig}}, \ and\ \bibinfo {author} {\bibfnamefont
  {T.}~\bibnamefont {Neupert}},\ }\href {\doibase 10.1038/s41567-018-0224-7}
  {\bibfield  {journal} {\bibinfo  {journal} {Nature Physics}\ }\textbf
  {\bibinfo {volume} {14}},\ \bibinfo {pages} {918} (\bibinfo {year}
  {2018})}\BibitemShut {NoStop}%
\bibitem [{\citenamefont {Aggarwal}\ \emph {et~al.}(2021)\citenamefont
  {Aggarwal}, \citenamefont {Zhu}, \citenamefont {Hughes},\ and\ \citenamefont
  {Madhavan}}]{Aggarwal-21}%
  \BibitemOpen
  \bibfield  {author} {\bibinfo {author} {\bibfnamefont {L.}~\bibnamefont
  {Aggarwal}}, \bibinfo {author} {\bibfnamefont {P.}~\bibnamefont {Zhu}},
  \bibinfo {author} {\bibfnamefont {T.~L.}\ \bibnamefont {Hughes}}, \ and\
  \bibinfo {author} {\bibfnamefont {V.}~\bibnamefont {Madhavan}},\ }\href
  {\doibase 10.1038/s41467-021-24683-8} {\bibfield  {journal} {\bibinfo
  {journal} {Nature Communications}\ }\textbf {\bibinfo {volume} {12}},\
  \bibinfo {pages} {4420} (\bibinfo {year} {2021})}\BibitemShut {NoStop}%
\bibitem [{\citenamefont {Drozdov}\ \emph {et~al.}(2014)\citenamefont
  {Drozdov}, \citenamefont {Alexandradinata}, \citenamefont {Jeon},
  \citenamefont {Nadj-Perge}, \citenamefont {Ji}, \citenamefont {Cava},
  \citenamefont {Bernevig},\ and\ \citenamefont {Yazdani}}]{Drozdov-14}%
  \BibitemOpen
  \bibfield  {author} {\bibinfo {author} {\bibfnamefont {I.~K.}\ \bibnamefont
  {Drozdov}}, \bibinfo {author} {\bibfnamefont {A.}~\bibnamefont
  {Alexandradinata}}, \bibinfo {author} {\bibfnamefont {S.}~\bibnamefont
  {Jeon}}, \bibinfo {author} {\bibfnamefont {S.}~\bibnamefont {Nadj-Perge}},
  \bibinfo {author} {\bibfnamefont {H.}~\bibnamefont {Ji}}, \bibinfo {author}
  {\bibfnamefont {R.~J.}\ \bibnamefont {Cava}}, \bibinfo {author}
  {\bibfnamefont {B.~A.}\ \bibnamefont {Bernevig}}, \ and\ \bibinfo {author}
  {\bibfnamefont {A.}~\bibnamefont {Yazdani}},\ }\href {\doibase
  10.1038/NPHYS3048} {\bibfield  {journal} {\bibinfo  {journal} {Nature
  Physics}\ }\textbf {\bibinfo {volume} {10}},\ \bibinfo {pages} {664}
  (\bibinfo {year} {2014})}\BibitemShut {NoStop}%
\bibitem [{\citenamefont {Wang}\ \emph {et~al.}(2019)\citenamefont {Wang},
  \citenamefont {Wieder}, \citenamefont {Li}, \citenamefont {Yan},\ and\
  \citenamefont {Bernevig}}]{Wang-19TMD}%
  \BibitemOpen
  \bibfield  {author} {\bibinfo {author} {\bibfnamefont {Z.}~\bibnamefont
  {Wang}}, \bibinfo {author} {\bibfnamefont {B.~J.}\ \bibnamefont {Wieder}},
  \bibinfo {author} {\bibfnamefont {J.}~\bibnamefont {Li}}, \bibinfo {author}
  {\bibfnamefont {B.}~\bibnamefont {Yan}}, \ and\ \bibinfo {author}
  {\bibfnamefont {B.~A.}\ \bibnamefont {Bernevig}},\ }\href {\doibase
  10.1103/PhysRevLett.123.186401} {\bibfield  {journal} {\bibinfo  {journal}
  {Phys.\ Rev.\ Lett.}\ }\textbf {\bibinfo {volume} {123}},\ \bibinfo {pages}
  {186401} (\bibinfo {year} {2019})}\BibitemShut {NoStop}%
\bibitem [{\citenamefont {Ezawa}(2019)}]{Ezawa-19TMD}%
  \BibitemOpen
  \bibfield  {author} {\bibinfo {author} {\bibfnamefont {M.}~\bibnamefont
  {Ezawa}},\ }\href {\doibase 10.1038/s41598-019-41746-5} {\bibfield  {journal}
  {\bibinfo  {journal} {Sci. Rep.}\ }\textbf {\bibinfo {volume} {9}},\ \bibinfo
  {pages} {5286} (\bibinfo {year} {2019})}\BibitemShut {NoStop}%
\bibitem [{\citenamefont {Qian}\ \emph {et~al.}(2022)\citenamefont {Qian},
  \citenamefont {Liu}, \citenamefont {Liu},\ and\ \citenamefont
  {Yao}}]{Qian-22}%
  \BibitemOpen
  \bibfield  {author} {\bibinfo {author} {\bibfnamefont {S.}~\bibnamefont
  {Qian}}, \bibinfo {author} {\bibfnamefont {G.-B.}\ \bibnamefont {Liu}},
  \bibinfo {author} {\bibfnamefont {C.-C.}\ \bibnamefont {Liu}}, \ and\
  \bibinfo {author} {\bibfnamefont {Y.}~\bibnamefont {Yao}},\ }\href {\doibase
  10.1103/PhysRevB.105.045417} {\bibfield  {journal} {\bibinfo  {journal}
  {Phys.\ Rev.\ B}\ }\textbf {\bibinfo {volume} {105}},\ \bibinfo {pages}
  {045417} (\bibinfo {year} {2022})}\BibitemShut {NoStop}%
\bibitem [{\citenamefont {Wrasse}\ and\ \citenamefont
  {Schmidt}(2014)}]{Wrasse-14}%
  \BibitemOpen
  \bibfield  {author} {\bibinfo {author} {\bibfnamefont {E.~O.}\ \bibnamefont
  {Wrasse}}\ and\ \bibinfo {author} {\bibfnamefont {T.~M.}\ \bibnamefont
  {Schmidt}},\ }\href {\doibase 10.1021/nl502481f} {\bibfield  {journal}
  {\bibinfo  {journal} {Nano Lett.}\ }\textbf {\bibinfo {volume} {14}},\
  \bibinfo {pages} {5717} (\bibinfo {year} {2014})}\BibitemShut {NoStop}%
\bibitem [{\citenamefont {Liu}\ \emph {et~al.}(2015)\citenamefont {Liu},
  \citenamefont {Qian},\ and\ \citenamefont {Fu}}]{Liu-15}%
  \BibitemOpen
  \bibfield  {author} {\bibinfo {author} {\bibfnamefont {J.~W.}\ \bibnamefont
  {Liu}}, \bibinfo {author} {\bibfnamefont {X.~F.}\ \bibnamefont {Qian}}, \
  and\ \bibinfo {author} {\bibfnamefont {L.}~\bibnamefont {Fu}},\ }\href
  {\doibase 10.1021/acs.nanolett.5b00308} {\bibfield  {journal} {\bibinfo
  {journal} {Nano Lett.}\ }\textbf {\bibinfo {volume} {15}},\ \bibinfo {pages}
  {2657} (\bibinfo {year} {2015})}\BibitemShut {NoStop}%
\bibitem [{\citenamefont {Sante}\ \emph {et~al.}(2017)\citenamefont {Sante},
  \citenamefont {Das}, \citenamefont {Bigi}, \citenamefont {Ergonenc},
  \citenamefont {Gurtler}, \citenamefont {Krieger}, \citenamefont {Schmitt},
  \citenamefont {Ali}, \citenamefont {Rossi}, \citenamefont {Thomale},
  \citenamefont {Franchini}, \citenamefont {Picozzi}, \citenamefont {Fujii},
  \citenamefont {Strocov}, \citenamefont {Sangiovanni}, \citenamefont
  {Vobornik}, \citenamefont {Cava},\ and\ \citenamefont
  {Panaccione}}]{Sante-17}%
  \BibitemOpen
  \bibfield  {author} {\bibinfo {author} {\bibfnamefont {D.~D.}\ \bibnamefont
  {Sante}}, \bibinfo {author} {\bibfnamefont {P.~K.}\ \bibnamefont {Das}},
  \bibinfo {author} {\bibfnamefont {C.}~\bibnamefont {Bigi}}, \bibinfo {author}
  {\bibfnamefont {Z.}~\bibnamefont {Ergonenc}}, \bibinfo {author}
  {\bibfnamefont {N.}~\bibnamefont {Gurtler}}, \bibinfo {author} {\bibfnamefont
  {J.~A.}\ \bibnamefont {Krieger}}, \bibinfo {author} {\bibfnamefont
  {T.}~\bibnamefont {Schmitt}}, \bibinfo {author} {\bibfnamefont {M.~N.}\
  \bibnamefont {Ali}}, \bibinfo {author} {\bibfnamefont {G.}~\bibnamefont
  {Rossi}}, \bibinfo {author} {\bibfnamefont {R.}~\bibnamefont {Thomale}},
  \bibinfo {author} {\bibfnamefont {C.}~\bibnamefont {Franchini}}, \bibinfo
  {author} {\bibfnamefont {S.}~\bibnamefont {Picozzi}}, \bibinfo {author}
  {\bibfnamefont {J.}~\bibnamefont {Fujii}}, \bibinfo {author} {\bibfnamefont
  {V.~N.}\ \bibnamefont {Strocov}}, \bibinfo {author} {\bibfnamefont
  {G.}~\bibnamefont {Sangiovanni}}, \bibinfo {author} {\bibfnamefont
  {I.}~\bibnamefont {Vobornik}}, \bibinfo {author} {\bibfnamefont {R.~J.}\
  \bibnamefont {Cava}}, \ and\ \bibinfo {author} {\bibfnamefont
  {G.}~\bibnamefont {Panaccione}},\ }\href {\doibase
  10.1103/PhysRevLett.119.026403} {\bibfield  {journal} {\bibinfo  {journal}
  {Phys.\ Rev.\ Lett.}\ }\textbf {\bibinfo {volume} {119}},\ \bibinfo {pages}
  {026403} (\bibinfo {year} {2017})}\BibitemShut {NoStop}%
\bibitem [{\citenamefont {Sihi}\ and\ \citenamefont
  {Pandey}(2021{\natexlab{a}})}]{Sihi-21}%
  \BibitemOpen
  \bibfield  {author} {\bibinfo {author} {\bibfnamefont {A.}~\bibnamefont
  {Sihi}}\ and\ \bibinfo {author} {\bibfnamefont {S.~K.}\ \bibnamefont
  {Pandey}},\ }\href {\doibase 10.1088/1361-648X/abeca8} {\bibfield  {journal}
  {\bibinfo  {journal} {J. Phys.: Condens. Matter}\ }\textbf {\bibinfo {volume}
  {33}},\ \bibinfo {pages} {225505} (\bibinfo {year}
  {2021}{\natexlab{a}})}\BibitemShut {NoStop}%
\bibitem [{\citenamefont {Sihi}\ and\ \citenamefont
  {Pandey}(2021{\natexlab{b}})}]{Sihi-22}%
  \BibitemOpen
  \bibfield  {author} {\bibinfo {author} {\bibfnamefont {A.}~\bibnamefont
  {Sihi}}\ and\ \bibinfo {author} {\bibfnamefont {S.~K.}\ \bibnamefont
  {Pandey}},\ }\href {\doibase 10.1088/1361-648X/ac5f62} {\bibfield  {journal}
  {\bibinfo  {journal} {J. Phys.: Condens. Matter}\ }\textbf {\bibinfo {volume}
  {34}},\ \bibinfo {pages} {245501} (\bibinfo {year}
  {2021}{\natexlab{b}})}\BibitemShut {NoStop}%
\bibitem [{\citenamefont {Wang}\ \emph {et~al.}(2018)\citenamefont {Wang},
  \citenamefont {Liu}, \citenamefont {Lu},\ and\ \citenamefont
  {Zhang}}]{Wang-18}%
  \BibitemOpen
  \bibfield  {author} {\bibinfo {author} {\bibfnamefont {Q.}~\bibnamefont
  {Wang}}, \bibinfo {author} {\bibfnamefont {C.-C.}\ \bibnamefont {Liu}},
  \bibinfo {author} {\bibfnamefont {Y.-M.}\ \bibnamefont {Lu}}, \ and\ \bibinfo
  {author} {\bibfnamefont {F.}~\bibnamefont {Zhang}},\ }\href {\doibase
  10.1103/PhysRevLett.121.186801} {\bibfield  {journal} {\bibinfo  {journal}
  {Phys.\ Rev.\ Lett.}\ }\textbf {\bibinfo {volume} {121}},\ \bibinfo {pages}
  {186801} (\bibinfo {year} {2018})}\BibitemShut {NoStop}%
\bibitem [{\citenamefont {Rachel}(2018)}]{rachel-18}%
  \BibitemOpen
  \bibfield  {author} {\bibinfo {author} {\bibfnamefont {S.}~\bibnamefont
  {Rachel}},\ }\href {\doibase 10.1088/1361-6633/aad6a6} {\bibfield  {journal}
  {\bibinfo  {journal} {Rep. Prog. Phys.}\ }\textbf {\bibinfo {volume} {81}},\
  \bibinfo {pages} {116501} (\bibinfo {year} {2018})}\BibitemShut {NoStop}%
\bibitem [{\citenamefont {Zlotnikov}\ \emph {et~al.}(2020)\citenamefont
  {Zlotnikov}, \citenamefont {Aksenov},\ and\ \citenamefont
  {Shustin}}]{zlotnikov-20}%
  \BibitemOpen
  \bibfield  {author} {\bibinfo {author} {\bibfnamefont {A.~O.}\ \bibnamefont
  {Zlotnikov}}, \bibinfo {author} {\bibfnamefont {S.~V.}\ \bibnamefont
  {Aksenov}}, \ and\ \bibinfo {author} {\bibfnamefont {M.~S.}\ \bibnamefont
  {Shustin}},\ }\href {\doibase 10.1134/S1063783420090371} {\bibfield
  {journal} {\bibinfo  {journal} {Phys. Solid State}\ }\textbf {\bibinfo
  {volume} {62}},\ \bibinfo {pages} {1612} (\bibinfo {year}
  {2020})}\BibitemShut {NoStop}%
\bibitem [{\citenamefont {Bir}\ and\ \citenamefont {Pikus}(1972)}]{bir-72}%
  \BibitemOpen
  \bibfield  {author} {\bibinfo {author} {\bibfnamefont {G.~L.}\ \bibnamefont
  {Bir}}\ and\ \bibinfo {author} {\bibfnamefont {G.~E.}\ \bibnamefont
  {Pikus}},\ }\href@noop {} {\emph {\bibinfo {title} {Symmetry and
  Strain-Induced Effects in Semiconductors}}}\ (\bibinfo  {publisher} {Nauka,
  Moscow},\ \bibinfo {year} {1972})\BibitemShut {NoStop}%
\bibitem [{\citenamefont {Hubbard}(1965)}]{hubbard-65}%
  \BibitemOpen
  \bibfield  {author} {\bibinfo {author} {\bibfnamefont {J.}~\bibnamefont
  {Hubbard}},\ }\href {\doibase https://doi.org/10.1098/rspa.1965.0124}
  {\bibfield  {journal} {\bibinfo  {journal} {Proc.\,Roy.\,Soc.\,A}\ }\textbf
  {\bibinfo {volume} {285}},\ \bibinfo {pages} {542} (\bibinfo {year}
  {1965})}\BibitemShut {NoStop}%
\bibitem [{\citenamefont {Ovchinnikov}\ and\ \citenamefont
  {Valkov}(2004)}]{ovchinnikov-04}%
  \BibitemOpen
  \bibfield  {author} {\bibinfo {author} {\bibfnamefont {S.~G.}\ \bibnamefont
  {Ovchinnikov}}\ and\ \bibinfo {author} {\bibfnamefont {V.~V.}\ \bibnamefont
  {Valkov}},\ }\href@noop {} {\emph {\bibinfo {title} {Hubbard Operators in the
  Theory of Strongly Correlated Electrons}}}\ (\bibinfo  {publisher} {Imperial
  College Press, London},\ \bibinfo {year} {2004})\BibitemShut {NoStop}%
\bibitem [{\citenamefont {Dyson}(1956)}]{dyson-56}%
  \BibitemOpen
  \bibfield  {author} {\bibinfo {author} {\bibfnamefont {F.}~\bibnamefont
  {Dyson}},\ }\href {\doibase DOI:https://doi.org/10.1103/PhysRev.102.1217}
  {\bibfield  {journal} {\bibinfo  {journal} {Phys.\,Rev.}\ }\textbf {\bibinfo
  {volume} {102}},\ \bibinfo {pages} {1217} (\bibinfo {year}
  {1956})}\BibitemShut {NoStop}%
\bibitem [{\citenamefont {Ivanov}\ and\ \citenamefont
  {Zaitsev}(1988)}]{ivanov-88}%
  \BibitemOpen
  \bibfield  {author} {\bibinfo {author} {\bibfnamefont {V.~A.}\ \bibnamefont
  {Ivanov}}\ and\ \bibinfo {author} {\bibfnamefont {R.~O.}\ \bibnamefont
  {Zaitsev}},\ }\href {\doibase 10.1142/S0217979288000524} {\bibfield
  {journal} {\bibinfo  {journal} {International Journal of Modern Physics B}\
  }\textbf {\bibinfo {volume} {2}},\ \bibinfo {pages} {689} (\bibinfo {year}
  {1988})}\BibitemShut {NoStop}%
\bibitem [{\citenamefont {Yan}\ \emph {et~al.}(2018)\citenamefont {Yan},
  \citenamefont {Song},\ and\ \citenamefont {Wang}}]{Yan-18}%
  \BibitemOpen
  \bibfield  {author} {\bibinfo {author} {\bibfnamefont {Z.}~\bibnamefont
  {Yan}}, \bibinfo {author} {\bibfnamefont {F.}~\bibnamefont {Song}}, \ and\
  \bibinfo {author} {\bibfnamefont {Z.}~\bibnamefont {Wang}},\ }\href {\doibase
  10.1103/PhysRevLett.121.096803} {\bibfield  {journal} {\bibinfo  {journal}
  {Phys.\ Rev.\ Lett.}\ }\textbf {\bibinfo {volume} {121}},\ \bibinfo {pages}
  {096803} (\bibinfo {year} {2018})}\BibitemShut {NoStop}%
\bibitem [{\citenamefont {Ikegaya}\ \emph {et~al.}(2021)\citenamefont
  {Ikegaya}, \citenamefont {Rui}, \citenamefont {Manske},\ and\ \citenamefont
  {Schnyder}}]{Ikegaya-21}%
  \BibitemOpen
  \bibfield  {author} {\bibinfo {author} {\bibfnamefont {S.}~\bibnamefont
  {Ikegaya}}, \bibinfo {author} {\bibfnamefont {W.~B.}\ \bibnamefont {Rui}},
  \bibinfo {author} {\bibfnamefont {D.}~\bibnamefont {Manske}}, \ and\ \bibinfo
  {author} {\bibfnamefont {A.~P.}\ \bibnamefont {Schnyder}},\ }\href {\doibase
  10.1103/PhysRevResearch.3.023007} {\bibfield  {journal} {\bibinfo  {journal}
  {Phys.\ Rev.\ Research}\ }\textbf {\bibinfo {volume} {3}},\ \bibinfo {pages}
  {023007} (\bibinfo {year} {2021})}\BibitemShut {NoStop}%
\bibitem [{\citenamefont {Fedoseev}(2022)}]{Fedoseev-22}%
  \BibitemOpen
  \bibfield  {author} {\bibinfo {author} {\bibfnamefont {A.}~\bibnamefont
  {Fedoseev}},\ }\href {\doibase 10.1103/PhysRevB.105.155423} {\bibfield
  {journal} {\bibinfo  {journal} {Phys.\ Rev.\ B}\ }\textbf {\bibinfo {volume}
  {105}},\ \bibinfo {pages} {155423} (\bibinfo {year} {2022})}\BibitemShut
  {NoStop}%
\end{thebibliography}%

\end{document}